\pdfoutput=1
\documentclass[PAPER, atlasdraft=false, cernpreprint, UKenglish, texmf, orcidlogo]{atlasdoc}
\usepackage{atlaspackage}
\usepackage{atlasbiblatex}

\usepackage{atlasphysics}
\graphicspath{{logos/}{figures/}}

\usepackage{ANA-HIGP-2024-23-PAPER-defs}
\usepackage{multirow}
\usepackage{booktabs}
\usepackage{makecell}

\AtlasTitle{Search for anomalies in vector-boson fusion production of the Higgs boson
in $H(\rightarrow \gamma\gamma) jj$ events using 164\,fb$^{-1}$ of $pp$ collision data
collected at $\sqrt{s}=13.6\,\text{TeV}$ with the ATLAS detector}

\AtlasAbstract{%
This article details two studies of Higgs boson properties using the vector-boson
fusion production mode and the $\gamma\gamma jj$ final state. Both efforts are based on a
data sample corresponding to 164\,fb$^{-1}$ of $\sqrt{s}=13.6\,\text{TeV}$ proton--proton collisions recorded
by the ATLAS experiment at the Large Hadron Collider. The first study employs matrix-element-based
optimal observables to constrain CP-odd couplings beyond the Standard Model within the Standard
Model Effective Field Theory framework, expressed in the Warsaw basis. The second study exploits angular
distributions to probe the Higgs boson's couplings to longitudinally and transversely polarised $W$ and $Z$
bosons in the production of the Higgs boson. To maximise the sensitivity, the constraints of the CP-odd
couplings are combined with those from a previous analysis performed in $\gamma\gamma jj$ events in a data sample
of proton--proton collisions at $\sqrt{s}=13\,\text{TeV}$, corresponding to an integrated luminosity of 140\,fb$^{-1}$.
A significant improvement with respect to the previous analysis is achieved through the implementation of a new
neural-network-based classification algorithm.
All measurements are in agreement with the Standard Model prediction of a CP-even Higgs boson with the expected
relative coupling strengths to longitudinally and transversely polarised vector bosons.
}

\AtlasRefCode{HIGP-2024-23}

\PreprintIdNumber{CERN-EP-2026-075}

\AtlasJournalRef{JHEP 09 (2026) 209}
\AtlasDOI{10.1007/JHEP09(2026)209}

\AtlasCoverEgroupAnalysisTeam{atlas-higp-2024-23-analysis-team}

\hypersetup{pdftitle={ATLAS document},pdfauthor={The ATLAS Collaboration}}

\begin{document}

\maketitle

\tableofcontents

\section{Introduction}\label{sec:intro}


The observation of the Higgs boson at the Large Hadron Collider (LHC) by the ATLAS and CMS
collaborations~\cite{HIGG-2012-27,CMS-HIG-12-028} was a major success.  Subsequent measurements have determined
its spin, parity, mass, width, and couplings to gauge bosons and fermions with increasing
precision~\cite{HIGG-2013-17,HIGG-2022-20,HIGP-2024-14,CMS-HIG-21-013,HIGG-2021-23,CMS-HIG-22-001}.  These precise measurements make the Higgs boson an indispensable tool for
testing the Standard Model (SM) and searching for physics beyond the Standard Model (BSM). In particular, the tensor
structure of the Higgs boson's couplings to vector bosons $HVV$ (where $V = W, Z$) is a sensitive probe for BSM
effects~\cite{Anderson:2013afp,Giudice:2007fh}, which motivates the search for anomalous contributions to the $HVV$ vertex.

\begin{figure}[!h]
\centering
\subfloat[]{\includegraphics[width=0.45\textwidth]{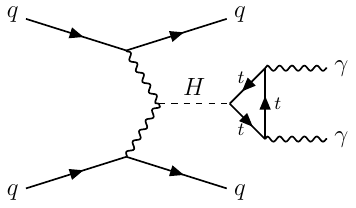}}
\subfloat[]{\includegraphics[width=0.45\textwidth]{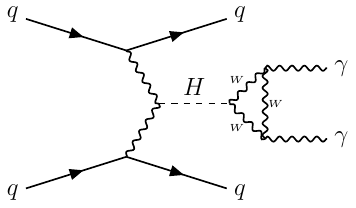}}
\caption{Representative lowest-order Feynman diagrams of vector-boson fusion production of a
Higgs boson and subsequent decay into two photons $H\rightarrow \gamma\gamma$ via (a) a top-quark loop and (b) a $W$-boson loop.}
\label{fig:FeynmanDiagram}
\end{figure}

This article presents results from two studies of the $HVV$ vertex in the vector-boson fusion (VBF) production of the Higgs boson,
using $H(\rightarrow \gamma\gamma) jj$ events. Representative leading-order
diagrams for the VBF production mode of the Higgs boson and its subsequent decay into two photons
are depicted in Figure~\ref{fig:FeynmanDiagram}. The first study searches for the violation of combined charge conjugation
and parity (CP) invariance in the Higgs-boson coupling to vector bosons. The second study probes the Higgs boson
couplings to longitudinally and transversely polarised vector bosons. These studies are based on
proton--proton ($pp$) collision data corresponding to an integrated luminosity of 164 fb$^{-1}$ collected with
the ATLAS detector at $\sqrt{s}=13.6\,\TeV$ from 2022 to 2024.

Searches for CP-violating effects in Higgs-boson couplings are well motivated because the known sources of
CP violation in nature are insufficient to explain the observed matter-antimatter asymmetry in our Universe
through baryogenesis~\cite{Sakharov_1991}. This strongly motivates further exploration, particularly as
extended scalar sector models can naturally introduce new sources of CP violation~\cite{Branco:2011iw}. At the
same time, studies of Higgs-boson couplings to longitudinally polarised vector bosons are also of great
importance. This is because the longitudinal polarisation states originate from the massless degrees of
freedom arising from the Higgs field during electroweak symmetry breaking. Within the SM, the Higgs boson's coupling
to these longitudinal states prevents the divergence of tree-level vector-boson scattering amplitudes at
high energies~\cite{Lee:1977yc}. If the Higgs boson were not a fundamental particle but a bound state, as
predicted in models with a composite Higgs boson~\cite{Kaplan:1983fs,Kaplan:1983sm}, the Higgs-boson couplings
to longitudinally and transversely polarised vector bosons may significantly deviate from their SM values.

Since the observation of the Higgs boson, the ATLAS and CMS collaborations have performed extensive searches for CP-violating effects in its couplings.
During Run 1 of the LHC, these efforts primarily focused on Higgs boson decays into $WW^*$ and $ZZ^*$~\cite{HIGG-2013-17}. Using 25\,fb$^{-1}$
of $pp$ collision data collected at centre-of-mass energies of $\sqrt{s}=7\,\TeV$ and $8\,\TeV$, both experiments provided the first
limits on CP-odd contributions to the $HVV$ vertex. With the same data sample, the first searches for CP-odd contributions in
the Higgs boson production were performed, targeting the VBF process in the $H \rightarrow \tau\tau$ decay channel~\cite{HIGG-2015-06}
and associated $VH$ production in the $H \rightarrow b\bar{b}$ decay channel~\cite{CMS-HIG-14-035}. The LHC experiments later expanded these searches to various
production processes, using the 140\,fb$^{-1}$ $pp$ data sample of Run 2 at $\sqrt{s}=13,\TeV$, including gluon--gluon fusion (ggF) in events with two associated
jets~\cite{HIGG-2017-13,CMS-HIG-20-007}, as well as the VBF~\cite{CMS-HIG-20-007,HIGG-2018-30,HIGG-2020-08,HIGG-2018-14,HIGP-2024-09,HIGP-2024-07}
and $t\bar{t}H$~\cite{HIGG-2019-01,HIGG-2020-03,CMS-HIG-21-006,CMS-HIG-19-013} production modes. These studies used events with Higgs boson decays into
$WW^{*}$, $\gamma\gamma$, $ZZ^{*}$, and $\tau\tau$. Building on the individual analyses, a combination of measurements of CP properties of Higgs boson
interactions to vector bosons was also performed~\cite{HIGP-2024-26}. In addition, searches for CP violation in the interaction between the Higgs boson and
$\tau$-leptons were performed by probing the $\tau$-lepton decay plane~\cite{HIGG-2019-10,CMS-HIG-20-006}.
To date, studies of Higgs-boson couplings to longitudinally and transversely polarised vector bosons have been
performed only in $H(\rightarrow WW^{*}\rightarrow e\nu\mu\nu) jj$ events~\cite{HIGG-2017-13}.

The two studies presented in this article probe CP- and polarisation-dependent observables constructed from the final-state particles in selected $\gamma\gamma jj$
events to test for deviations from the SM expectations. Analysis regions are defined based on requirements on the kinematic properties of the final-state
particles and event-level quantities. To enhance the separation of the VBF signal from the background processes, a multi-class classifier based on a neural
network is employed. The primary backgrounds to the VBF $H(\rightarrow \gamma\gamma)$ signal consist of final states with two photons and multiple jets
and events where one or more jets are misidentified as photons. These backgrounds are determined using data-driven methods. Subleading backgrounds from other Higgs
boson production modes are modelled via simulation, with their normalisations either constrained by a profile-likelihood fit to the data (in the case of the ggF production
mode) or fixed to the Standard Model predictions. The results of the CP studies are interpreted within
the Standard Model Effective Field Theory (SMEFT) framework~\cite{Brivio:2017vri} using the Warsaw basis~\cite{Buchmuller:1985jz,WarsawBasis}.
Complementing these CP studies, the results from the polarisation studies are interpreted as polarisation-dependent coupling scale factors between the Higgs boson and
the vector bosons~\cite{Brehmer:2014pka}.

To maximise the sensitivity, the results of the CP study are combined with a previous analysis performed in $\gamma\gamma jj$ events in a data sample of proton--proton
collisions at $\sqrt{s}=13\,\TeV$, corresponding to an integrated luminosity of 140\,fb$^{-1}$~\cite{HIGG-2020-08}. Beyond the increased statistics provided by the larger
data sample, the analysis presented in this article introduces several methodological improvements over the previous Run-2 result, most notably the implementation of more
sophisticated machine learning techniques for signal and background classification.
Furthermore, the polarisation studies, performed here for the first time in the VBF $H(\rightarrow \gamma\gamma)$ channel, benefit significantly from the clean experimental
signature and manageable background composition of the diphoton plus jets final state. In contrast, prior
measurements in $H(\rightarrow WW^{*}\rightarrow e\nu\mu\nu)$ events were limited by the modelling of the
complex background composition and the associated uncertainties, despite the larger Higgs boson branching
fraction.


%

%
\section{Theoretical framework and analysis methodology}\label{sec:Observables}

\subsection{CP Violation in the EFT Framework}

For the studies testing the CP-invariance of the Higgs boson coupling to vector bosons, the SMEFT
framework is used to parameterise possible deviations from the SM\@. The SMEFT framework supplements
the SM Lagrangian with operators $\mathcal{O}_i$ of mass-dimension $D > 4$ that are invariant under the
SM gauge group $SU(3)_C\times SU(2)_L\times U(1)_Y$. The resulting effective Lagrangian is expressed as:

\begin{equation}
\mathcal{L}_\mathrm{SMEFT} = \mathcal{L}_{SM} + \sum_i \frac{c^{(D)}_i}{\Lambda^{D-4}} \mathcal{O}^{(D)}_i
\end{equation}

where $\Lambda$ is the energy scale of new physics, and the dimensionless Wilson coefficients $c_i$ govern the strength of
the new interactions. Since the dimension-5 operator is typically ignored as it violates lepton number conservation~\cite{Weinberg:1979sa},
dimension-6 operators are expected to provide the leading contributions to physical observables.

Among the complete sets of dimension-6 operators, the Warsaw basis is the most widely adopted for LHC
results~\cite{deFlorian:2016spz}. In this basis, the three CP-odd operators $\mathcal{O}_{\Phi \widetilde{W}}$,
$\mathcal{O}_{\Phi\widetilde{W}B}$, and $\mathcal{O}_{\Phi \widetilde{B}}$, along with their corresponding Wilson coefficients
$c_{H\widetilde{W}}$, $c_{H\widetilde{W}B}$, and $c_{H\widetilde{B}}$, contribute to the $HVV$ vertex.
In the VBF $H \rightarrow \gamma\gamma$  process, these couplings affect both the production vertex and the loop-induced decay. While all three
interactions modify the Higgs boson branching ratio, the $H \rightarrow \gamma\gamma$ decay itself remains insensitive to CP-violating effects.
This is because in two-body decays of a spin-0 particle, all kinematic observables are strictly determined by energy-momentum conservation.
Consequently, the decay kinematics lack the degrees of freedom required to construct CP-sensitive observables. In contrast, the production vertex
offers significant sensitivity. Given that VBF production is dominated by $W^+ W^- \rightarrow H$ contributions, this analysis is primarily sensitive
to $\mathcal{O}_{\Phi \widetilde{W}}$ and focuses on constraining $c_{H\widetilde{W}}$, while all other Wilson coefficients affecting the
$HVV$ vertex are assumed to be zero.

\subsection{Polarisation-dependent coupling-strength scale factors}

To investigate the Higgs boson couplings to longitudinally and transversely polarised $W$ and $Z$ bosons, polarisation-dependent coupling-strength scale
factors are introduced. Following the framework established in Ref.~\cite{Brehmer:2014pka}, these factors are defined as:

\begin{equation}
a_\mathrm{L} =  \frac{g_{\scriptscriptstyle HV_\mathrm{L}V_\mathrm{L}}}{g_{\scriptscriptstyle HVV}},\,\,\, a_\mathrm{T} =  \frac{g_{\scriptscriptstyle HV_\mathrm{T}V_\mathrm{T}}}{g_{\scriptscriptstyle HVV}}
\end{equation}

where $g_{\scriptscriptstyle HVV}$ denotes the SM $HVV$ coupling strength, which is independent of the polarisation state, while $g_{\scriptscriptstyle HV_\mathrm{L}V_\mathrm{L}}$ and
$g_{\scriptscriptstyle HV_\mathrm{T}V_\mathrm{T}}$ represent the measured polarisation-dependent couplings. In this study, the vector boson polarisations are defined in the Higgs boson
rest frame such that the mixed-polarisation couplings $g_{\scriptscriptstyle HV_\mathrm{L}V_\mathrm{T}}$ vanish and do not contribute to the production process. This parameterisation of
the coupling-strength focuses exclusively on these scale factors, and other BSM effects are not considered.
The Standard Model is characterised by $a_\mathrm{L}=a_\mathrm{T}=1$.

\subsection{CP- and polarisation-dependent observables}

Modifications to the $HVV$ vertex from higher-dimensional operators impact the event kinematics and the total cross-section of Higgs boson production via VBF\@.
The resulting squared matrix element, for a set of CP-odd couplings $c_i$, is then given by:

\begin{align}
\left|\mathcal{M}\right|^2  &= \left|\mathcal{M}_\mathrm{SM} + \sum_{i}\frac{c_i}{\Lambda^2} \mathcal{M}_{\mathrm{BSM},i}\right|^2  \\ \nonumber
&= \left|\mathcal{M}_{SM}\right|^2 + 2 \sum_i \frac{c_i}{\Lambda^2}\mathrm{Re}\left(\mathcal{M}^{*}_{\mathrm{SM}}\mathcal{M}_{\mathrm{BSM},i}\right) +\sum_i \sum_j \frac{c_i c_j}{\Lambda^4} \mathrm{Re}\left(\mathcal{M}^{*}_{\mathrm{BSM},i}\mathcal{M}_{\mathrm{BSM},j}\right)\,.
\end{align}

The first term in the expansion corresponds to the squared SM matrix element. The second term (referred to in the following as ``linear term'') is composed of the cross terms between the SM and BSM
couplings and represents the interference between the SM and BSM contributions. Finally, the third term accounts for the quadratic BSM contributions and the interference between
different BSM terms (referred to in the following as ``quadratic term''). The SM and quadratic terms are CP-even. In contrast, the linear term is CP-odd and provides a potential source
of CP-violating effects in the Higgs boson couplings. Because the linear term is CP-odd, its integral over a CP-symmetric phase space vanishes. Consequently, it does not contribute
to the total cross-section. While the contributions from the quadratic term increases the total cross-section, its effect is not exploited in the current analysis because the
observed event rate may be simultaneously influenced by contributions from BSM CP-even operators.

To maximise statistical sensitivity to CP-violating effects, an ``Optimal Observable'' ($\ensuremath{\mathcal{OO}}$)~\cite{Diehl:1997ft, Nachtmann:2004fy} is defined as
the ratio of the interference term to the SM-only squared matrix element. For a specific CP-odd coupling $c_{i}$, this is given by:

\begin{equation} \label{eq:OO1_defintion}
\ensuremath{\mathcal{OO}}_{i} = \frac{2\mathrm{Re}\left(\mathcal{M}^{*}_{\mathrm{SM}}\mathcal{M}_{\mathrm{BSM},i}\right)}{|\mathcal{M}_{\mathrm{SM}}|^2}\,.
\end{equation}

\begin{figure}[!h]
\centering
\subfloat[]{\includegraphics[width=0.495\textwidth]{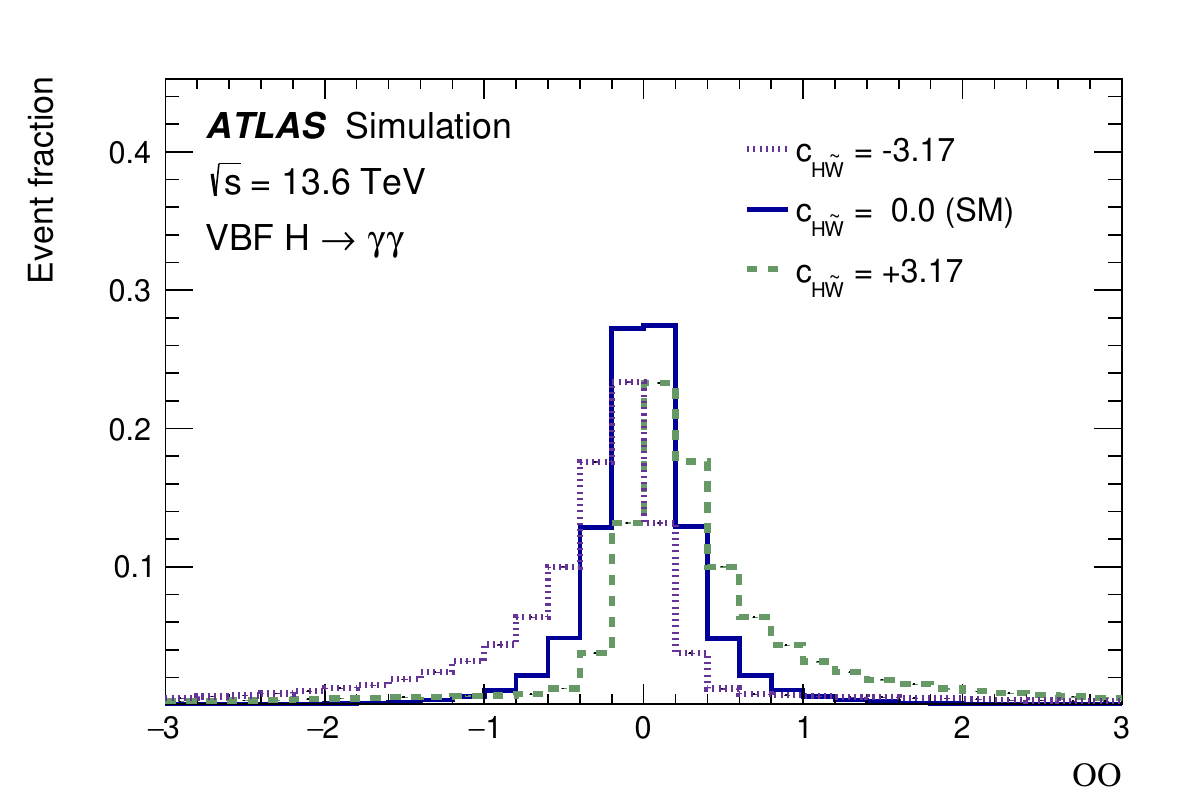}}
\subfloat[]{\includegraphics[width=0.495\textwidth]{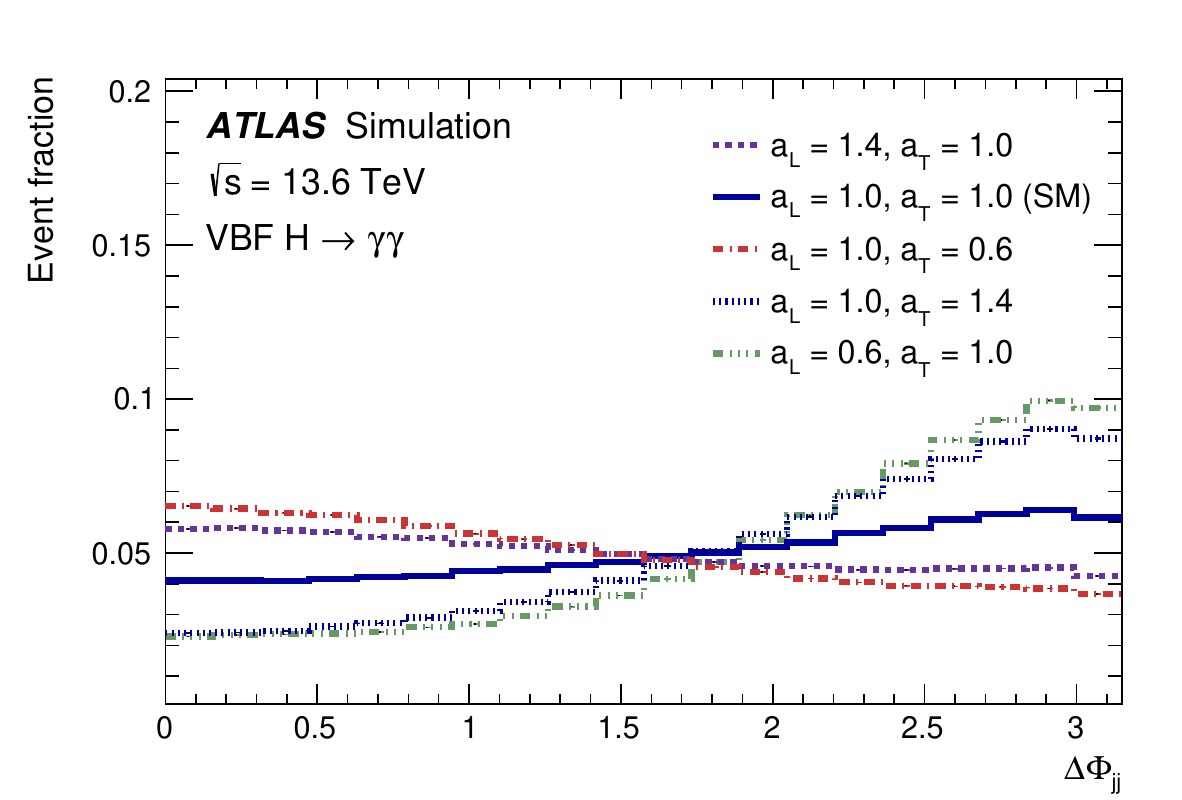}}
\caption{Distributions of (a) the $\ensuremath{\mathcal{OO}}$ observable shown for various configurations of the Wilson coefficient $\cHWtil$, and
(b) the $\DPhijj\,$ observable for various configurations of the $\aL$ and $\aT$ parameters. These comparisons are performed in VBF events
at the reconstruction level using the predictions of the \MGNLO + \PYTHIA8~\cite{Alwall:2014hca,Sjostrand:2014zea} generators.}
\label{fig:CPodd_and_pol_observables}
\end{figure}

By construction, these observables are CP-odd. In the absence of CP violation and assuming negligible contributions from rescattering~\cite{Brehmer:2017lrt}
(i.e. new particles being on the mass shell in loop corrections to the $HVV$ vertex), their distributions are symmetric around a vanishing mean. Consequently, any
observed asymmetry would provide direct evidence of CP violation in the $HVV$ vertex, a feature not shared by CP-even observables derived from squared matrix elements.
In this analysis, the $\ensuremath{\mathcal{OO}}$ observable is evaluated for the operator $\mathcal{O}_{\Phi \widetilde{W}}$ and the corresponding Wilson
coefficient $c_{H\widetilde{W}}$.
Figure~\ref{fig:CPodd_and_pol_observables} (a) illustrates the distribution of the $\ensuremath{\mathcal{OO}}$ for SM VBF Higgs boson production (where $\cHWtil = 0$) and
for two BSM scenarios where $\cHWtil \neq 0$. These distributions, shown after the selection requirements described in Section~\ref{sec:Strategy}, demonstrate
how the mean of the distribution shifts toward positive or negative values depending on the sign of the $\cHWtil$ parameter.\par

The matrix elements used in the construction of the $\ensuremath{\mathcal{OO}}$ are calculated at leading-order (LO) accuracy in QCD using
\MGNLO~\cite{Alwall:2014hca}, with the \NNPDF[3.0nlo]~\cite{Ball:2014uwa} set of parton distribution functions (PDFs). These matrix elements are
evaluated using the four-momenta of the Higgs boson, reconstructed from the two selected photons, and the two VBF-tagged jets.
At the reconstruction level, the momentum fractions of the initial-state partons are derived in the reference frame of the ATLAS detector as:
\begin{equation}
x^{\mathrm{reco}}_{1,2} = \frac{m_{Hjj} \cdot e^{\pm y_{Hjj}}}{\sqrt{s}}
\end{equation}
by exploiting energy and momentum conservation, where $m_{Hjj}$ and $y_{Hjj}$ are the invariant mass and rapidity of the Higgs boson + dijet system, and $\sqrt{s}$
is the centre-of-mass energy. A detailed description of this calculation can be found in Ref.~\cite{HIGG-2018-14}. \par

A simple polarisation-dependent observable, suggested in Ref.~\cite{Brehmer:2014pka} for constraining the coupling-strength scale factors
$a_\mathrm{L}$ and $a_\mathrm{T}$, is the absolute azimuthal angular difference between the two VBF-tagged jets ($\Delta\Phi_{jj}$). This variable is highly sensitive
to the polarisation states of the vector bosons in the VBF production mode. Consequently, $\Delta\Phi_{jj}$ is employed in this analysis
to provide sensitivity to these polarisation-dependent scale factors. Figure~\ref{fig:CPodd_and_pol_observables} (b) shows the distribution of the $\Delta\Phi_{jj}$
observable for SM VBF Higgs boson production (where $\aL = \,$\aT$ = 1$) and for four BSM scenarios where either $\aL$ or $\aT$ are varied.

%

%
\section{ATLAS detector}\label{sec:detector}


\newcommand{\AtlasCoordFootnote}{%
ATLAS uses a right-handed coordinate system with its origin at the nominal interaction point
(IP) in the centre of the detector and the \(z\)-axis along the beam pipe.
The \(x\)-axis points from the IP to the centre of the LHC ring,
and the \(y\)-axis points upwards. Polar coordinates \((r,\phi)\) are used in the transverse
plane, \(\phi\) being the azimuthal angle around the \(z\)-axis.
The pseudorapidity is defined in terms of the polar angle \(\theta\) as \(\eta = -\ln \tan(\theta/2)\)
and is equal to the rapidity $ y = \frac{1}{2} \ln \left( \frac{E + p_z}{E - p_z} \right) $
in the relativistic limit. Angular distance is measured in units
of \(\Delta R \equiv \sqrt{(\Delta y)^{2} + (\Delta\phi)^{2}}\).
The transverse energy is defined as $\ET = E/\cosh(\eta)$.
}

The ATLAS detector~\cite{PERF-2007-01,GENR-2019-02} at the LHC covers nearly the entire solid angle around the collision point.\footnote{\AtlasCoordFootnote}
It consists of an inner tracking detector surrounded by a thin superconducting solenoid, electromagnetic (EM) and hadronic calorimeters,
and a muon spectrometer incorporating three large superconducting air-core toroidal magnets.

The inner-detector system (ID) is immersed in a \qty{2}{\tesla} axial magnetic field
and provides charged-particle tracking in the range of \(|\eta| < 2.5\).
The high-granularity silicon pixel detector provides full azimuthal coverage around the interaction region
and typically provides four measurements per track,
the first hit generally being in the insertable B-layer (IBL).
It is followed by the SemiConductor Tracker (SCT), which usually provides eight measurements per track.
These silicon detectors are complemented by the transition radiation tracker (TRT),
which enables radially extended track reconstruction up to \(|\eta| = 2.0\).
The TRT also provides electron identification information
based on the fraction of hits (typically 30 in total) above a higher energy-deposit threshold corresponding to transition radiation.

The calorimeter system covers the pseudorapidity range \(|\eta| < 4.9\).
Within the region \(|\eta|< 3.2\), electromagnetic calorimetry is provided by barrel and
endcap high-granularity lead/liquid-argon (LAr) calorimeters,
with an additional thin LAr presampler covering \(|\eta| < 1.8\)
to correct for energy loss in material upstream of the calorimeters.
Hadronic calorimetry is provided by the steel/scintillator-tile calorimeter,
segmented into three barrel structures within \(|\eta| < 1.7\), and two copper/LAr hadronic endcap calorimeters.
The solid angle coverage is completed with forward copper/LAr and tungsten/LAr calorimeter modules
optimised for electromagnetic and hadronic energy measurements, respectively.

The muon spectrometer (MS) comprises separate trigger and
high-precision tracking chambers measuring the deflection of muons in a magnetic field generated by the superconducting air-core toroidal magnets.
The field integral of the toroids ranges between \num{2.0} and \qty{6.0}{\tesla\metre}
across most of the detector.
Three layers of precision chambers, each consisting of layers of monitored drift tubes, cover the region \(|\eta| < 2.7\),
except in the innermost layer of the endcap region, where layers of small-strip thin-gap chambers and Micromegas chambers both provide precision
tracking in the region \(1.3 < |\eta| < 2.7\). The muon trigger system covers the range \(|\eta| < 2.4\) with resistive-plate chambers in the
barrel, thin-gap chambers in the endcap regions, and the small-strip thin-gap chambers and Micromegas chambers in the innermost
layer of the endcap.

The luminosity is measured mainly by the LUCID--2 detector that records Cherenkov light produced in the quartz windows of photomultipliers located close to the beampipe.

Events were selected by the first-level trigger system implemented in custom hardware,
followed by selections made by algorithms implemented in software in the high-level trigger~\cite{TRIG-2022-01}.
The first-level trigger accepted events from the \qty{40}{\MHz} bunch crossings at a rate close to \qty{100}{\kHz},
which the high-level trigger further reduced to record complete events to disk at an average rate of about \qty{3}{\kHz}.

A software suite~\cite{SOFT-2022-02} is used in data simulation, in the reconstruction
and analysis of real and simulated data, in detector operations, and in the trigger and data acquisition
systems of the experiment.


%

%
\section{Data and simulated event samples}\label{sec:DataAndMC}

The $pp$ collision data at $\sqrt{s} = 13.6\,\TeV$ used in these analyses were recorded with the ATLAS detector between 2022 and 2024, corresponding
to a total integrated luminosity of $163.9 \pm 3.1$\,fb$^{-1}$~\cite{ATL-DAPR-PUB-2025-001}. The data are required to satisfy criteria ensuring
that the detector was in good operating condition~\cite{DAPR-2018-01}. Monte Carlo (MC) simulation samples were used to model the signal processes,
assist in background estimates, and derive modelling uncertainties. All MC simulation samples were processed using a hybrid approach:
the geometry and response of the ATLAS Inner Detector and muon spectrometer were simulated using \GEANT~\cite{Agostinelli:2002hh} (G4),
while the calorimeter response was modelled using AtlFast3 (AF3)~\cite{SIMU-2018-04}. AF3 combines parameterised approaches with
machine-learning techniques to simulate particle showers with a level of precision comparable to \GEANT, while being significantly faster.
For example, for single photons with an energy of $65\,\GeV$ and a pseudorapidity of $0.20 < |\eta| < 0.25$, AF3 is approximately
150 times faster than G4 when the calorimeter simulation alone is considered~\cite{SIMU-2018-04}.
The studies presented in this article are the first by the ATLAS Collaboration to exclusively use MC samples produced
with the AF3 detector simulation.

To study the CP properties of the Higgs boson couplings to vector bosons, the VBF production of the
Higgs boson and the subsequent decay of the Higgs boson into photons were simulated at LO accuracy
in QCD using the matrix element (ME) generator \MGNLO[3.3.1], the \NNPDF[3.0nlo] PDF set, and the
FeynRules model~\cite{Alloul:2013bka} SMEFTsim\_U35\_MwScheme~\cite{Brivio:2017btx, Brivio:2020onw}.
\PYTHIA[8.306]~\cite{Sjostrand:2014zea} with the A14 set of tuned parameters~\cite{ATL-PHYS-PUB-2014-021} was used to
model the parton shower, hadronisation, and the underlying event. While signal samples were generated for a discrete set of
$c_{H\widetilde{W}}$ values ($c_{H\widetilde{W}}\in\left\{-3.17, 0.0, 1.5, 3.17\right\}$ at $\Lambda = 1\,\TeV$),
parameter morphing~\cite{HIGG-2013-27,Baak:2014fta} was used to model the signal across a continuous range of coupling scenarios by
interpolating between the simulated basis points\footnote{The signal samples corresponding to $c_{H\widetilde{W}}\in\left\{-3.17, 0.0, 3.17\right\}$
were used to build the morphing basis, while the signal sample corresponding to $c_{H\widetilde{W}}=1.5$ was used to validate the morphing basis.}. These specific
$c_{H\widetilde{W}}$ values were chosen to maximise the statistical stability of the morphing procedure throughout the target parameter space.

For the studies of the Higgs boson couplings to longitudinally and transversely polarised $W$ and $Z$ bosons, the VBF production of the Higgs boson
and the subsequent decay into photons were simulated at LO in QCD using \MGNLO[2.8.3.2] and the \NNPDF[2.3nlo]~\cite{Ball:2012cx} PDF set.
The helicity amplitudes used in the ME generation of the Higgs boson production were modified to account for deviations in the
Higgs boson coupling strengths to the polarised vector bosons within the Higgs boson rest frame, following the prescriptions in Ref.~\cite{Brehmer:2014pka}.
Signal samples were produced for the following benchmark scenarios: $(\aL, \aT) \in \left\{(1.0,1.0), (1.4,1.0), (1.2,1.0), (1.0,1.4), (1.0,1.2), (1.0,0.6),
(0.6,1.0)\right\}$. Parton shower, hadronisation, and the underlying event were simulated using the \Pythia[8.308] generator with the A14 set of tuned parameters.

In both analyses, the diphoton background processes are categorised into two distinct components: resonant and non-resonant. The resonant background consists of
$\gamma\gamma$ production from other Higgs boson modes, including gluon--gluon fusion (ggF), associated production with a vector boson ($VH$), and production in
association with a pair of top- or bottom-quarks ($t\bar{t}H$ or $b\bar{b}H$) or a single top quark ($tHq$ and $tHW$). Conversely, the non-resonant component
includes continuum diphoton production in association with jets. Additional contributions from events where one or more jets are falsely
reconstructed as photons ($\gamma j$ or $jj$), are estimated by using a data-driven approach, as described in Section~\ref{sec:Modelling}.\par

The Higgs boson background processes (ggF, $VH$, $t\bar{t}H$ and $b\bar{b}H$) were simulated using the \POWHEGBOX[v2]~\cite{Frixione:2007nw,Nason:2004rx,Frixione:2007vw,Alioli:2010xd}
generator with the \PDFforLHC[21] PDF set~\cite{Ball:2022oua}. \Pythia[8.310] was subsequently employed to model the parton showering, hadronisation, and the underlying event.
Matrix elements for the ggF and $VH$ processes were calculated at next-to-leading-order (NLO) accuracy in QCD for diagrams with up to one extra parton following the MiNLO~\cite{Hamilton:2012np,Campbell:2012am,Hamilton:2012rf}
approach, though the $gg\rightarrow ZH$ process was computed only at LO accuracy. The $t\bar{t}H$ and $b\bar{b}H$ processes were also simulated at NLO accuracy in QCD\@. Finally, the production
of a Higgs boson in association with a single top quark ($tHq$ and $tHW$) was simulated using the \MGNLO generator and the \NNPDF[3.0nlo] PDF set to calculate the MEs at
NLO accuracy, while \Pythia with the A14 tune was used for the modelling of the parton shower, hadronisation, and underlying event. \par

The production of non-resonant diphoton + jets ($\gamma\gamma$+jets) events was simulated using the \Sherpa[2.2.14]~\cite{Bothmann:2019yzt} generator
with the \NNPDF[3.0nnlo]~\cite{Ball:2014uwa} PDF set. The ME was calculated based on the \textsc{Comix}~\cite{Gleisberg:2008fv} and \OPENLOOPS~\cite{Buccioni:2019sur,Cascioli:2011va,Denner:2016kdg}
libraries at NLO accuracy in QCD for diagrams with up to one additional parton emissions, and LO accuracy in QCD for diagrams with two or three additional
parton emissions. The \MEPSatNLO prescription~\cite{Hoeche:2011fd,Hoeche:2012yf,Catani:2001cc,Hoeche:2009rj} was used to merge the ME and the \Sherpa
parton shower~\cite{Schumann:2007mg}, which is based on a set of tuned parameters developed by the \Sherpa authors. A dynamic merging cut~\cite{Siegert:2016bre} of
10\,\GeV was used and photons were required to be isolated according to a smooth-cone isolation criterion~\cite{Frixione:1998jh}.

All simulated events were generated at a centre-of-mass energy $\sqrt{s}$ of $13.6~\TeV$ and were normalised to the state-of-the-art cross-section predictions\footnote{The VBF BSM signal
hypotheses were normalised by applying a global correction factor, defined as the ratio of the higher-order cross-section to the
leading-order cross section of the SM hypothesis to ensure a consistent treatment of higher-order effects across all coupling scenarios.}~\cite{deFlorian:2016spz}.
Simulated events containing Higgs bosons were generated with a Higgs boson mass of 125.0\,\GeV, a Higgs boson width of 4.07\,\MeV, and a branching ratio of
$(2.27 \pm 0.07) \times 10^{-3}$ for the $H\rightarrow \gamma\gamma$ decay~\cite{deFlorian:2016spz}. Furthermore, \textsc{EvtGen}\,2.1.1~\cite{Lange:2001uf}
was used for all event samples except for the ones produced with \Sherpa to unify the properties of the bottom and charm hadron decays.

The effect of multiple interactions in the same and neighbouring bunch crossings (\pileup)
was modelled by overlaying~\cite{SIMU-2020-01} the simulated hard-scattering event
with inelastic $pp$ events generated from a mix of
\EPOS[2.0.1.4]~\cite{Werner:2005jf} and \PYTHIA[8.308]~\cite{Bierlich:2022pfr}.
The \EPOS events were generated with the \EPOS LHC tune~\cite{Pierog:2013ria}
and the \PYTHIA events with the A3 tune~\cite{ATL-PHYS-PUB-2016-017}
and the \NNPDF[2.3lo]~\cite{Ball:2012cx} set of PDFs. \PYTHIA pileup events include
either a high transverse momentum (\pt) jet, a prompt photon, or a lepton from a $b$-hadron decay,
while \EPOS was filtered to simulate all remaining pileup events in the overlay sample.
The individual simulations were first reweighted to ensure a smooth connection across jet \pt then
the combination reweighted to match the distribution of the actual number of interactions per bunch
crossing (\(\mu\)) measured in data\footnote{The average (most probable) number of interactions per bunch
crossing were 42 (50) during the 2022 run, 51 (58) in the 2023 run, and 54 (63) in the 2024 run.}
during the 2022, 2023, and 2024 runs of the LHC\@. A summary of all Monte Carlo samples used in the
two analyses are presented in Table~\ref{tab:generators}.

\begin{table}[h!]
\centering
\renewcommand{\arraystretch}{1.1}
\caption{Summary of the Monte Carlo generators employed to produce the various signal and background
processes used for this analysis. The perturbative accuracy (in QCD and, if relevant, in EW corrections)
of the total production cross section $\sigma_\mathrm{prod}$ is stated for each process~\cite{deFlorian:2016spz}. The
order at which the corresponding matrix elements are calculated in the Monte Carlo simulation is not
necessarily the same as in the cross-section calculation.}
\resizebox{1.0\textwidth}{!}{
\begin{tabular}{c c c}
\toprule
Process                                      & Generator                        &  Perturbative accuracy of $\sigma_\mathrm{prod}$  \\
\toprule
VBF $H\rightarrow \gamma\gamma$ (\cHWtil variations)                             & \MGNLO[3.3.1]   + \PYTHIA[8.306]   & NNLO (QCD) + NLO (EW)   \\
VBF $H\rightarrow \gamma\gamma$ ($a_\mathrm{L}$ and $a_\mathrm{T}$ variations)   & \MGNLO[2.8.3.2] + \PYTHIA[8.308]   & NNLO (QCD) + NLO (EW)   \\
\midrule
$gg \rightarrow H\rightarrow \gamma\gamma$                                       & \POWHEGBOX[v2] + \Pythia[8.310]    &  NNNLO (QCD) + NLO (EW) \\
$VH(\rightarrow \gamma\gamma)$                                                   & \POWHEGBOX[v2] + \Pythia[8.310]    &  NNLO (QCD) + NLO (EW)  \\
$b\bar{b}H(\rightarrow \gamma\gamma)$                                            & \POWHEGBOX[v2] + \Pythia[8.310]    &  NLO                    \\
$t\bar{t}H(\rightarrow \gamma\gamma)$                                            & \POWHEGBOX[v2] + \Pythia[8.310]    &  NLO (QCD) + NLO (EW)   \\
$tHq$                                                                            & \MGNLO[3.5.1]  + \Pythia[8.309]    &  NLO                    \\
$tHW$                                                                            & \MGNLO[3.5.3]  + \Pythia[8.310]    &  NLO                    \\
\midrule
$pp\rightarrow \gamma\gamma$                                                     &         \Sherpa[2.2.14]            &  NLO                    \\
\bottomrule
\end{tabular}}
\label{tab:generators}
\end{table}


%

%
\section{Event reconstruction}\label{sec:Reco}


Charged-particle tracks are reconstructed in the ID using either the standard reconstruction algorithm~\cite{PERF-2015-08}
or an alternative procedure that accounts for bremsstrahlung energy losses during pattern recognition~\cite{PERF-2017-01}.
Tracks matched to EM calorimeter clusters~\cite{ATL-LARG-PUB-2008-002} are refitted using a Gaussian Sum Filter (GSF) algorithm
to improve the track parameter estimate. These GSF tracks are used for the reconstruction of converted photons,
while standard tracks are used for the reconstruction of collision vertices~\cite{ATL-PHYS-PUB-2015-026}.

Photons are reconstructed from variable-size topological clusters~\cite{PERF-2014-07} built from electromagnetic
calorimeter cells with energy deposits significantly above the noise threshold~\cite{PERF-2017-01}. Photon candidates
are classified as converted if matched to either two tracks forming a conversion vertex in the silicon detectors
or a single track originating from the silicon detectors consistent with an electron but lacking hits in the
innermost pixel layer. Otherwise, they are classified as unconverted. All photon candidates must have
$\ET > 25\,\GeV$ and fall within the acceptance region of the finely segmented first layer of the EM
calorimeter ($|\eta|<2.37$), excluding the transition region between the barrel and end cap calorimeters,
$1.37 < |\eta|<1.52$. Their energy measurement is calibrated using methods described in
Ref.~\cite{EGAM-2021-02}, including dedicated corrections for samples simulated with AF3.

Photon candidates must fulfil the identification criteria of the \textit{Tight} operating point ~\cite{EGAM-2021-01}, which are
based on the shape of the electromagnetic shower in the EM calorimeter and the leakage of energy into the hadronic calorimeter.
To suppress non-prompt photons from hadron decays, candidates must satisfy both calorimeter- and track-based isolation requirements.
The calorimeter-based isolation variable is defined as the sum of cluster transverse energies in a cone of $\Delta R = 0.2$ around the photon
candidate, with corrections for energy leakage, pile-up, and the underlying event. This variable is required
to be less than $6.5\%$ of the photon transverse energy. Similarly, the track-based isolation requires the scalar sum of $\pT$ for all
tracks within a cone of $\Delta R = 0.2$ around the photon to be less than $5\%$ of the photon's transverse energy, selecting only tracks
with $\pT > 1\,\GeV$ that are matched to the primary vertex and not associated with a conversion vertex.

Jets are reconstructed using the anti-$k_{t}$ algorithm~\cite{Cacciari:2008gp} as implemented in the
\textsc{FastJet} package~\cite{Fastjet}, with a radius parameter $R=0.4$ and particle-flow objects as
input~\cite{PERF-2015-09}. Jets are required to have $p_{\text{T}}>20\,\GeV$ and $|\eta|<4.4$. To
mitigate pile-up contamination, two multivariate classifiers are applied: one based on calorimeter and
tracking information for jets with $\pT < 60\,\GeV$ and $|\eta| < 2.5$~\cite{PERF-2014-03},
and another based on jet shapes and topological correlations for jets with $\pT < 60\,\GeV$
and $|\eta| > 2.5$~\cite{ATL-PHYS-PUB-2019-026}. Jet momenta are corrected for passive material losses,
non-compensating calorimeter response, and average pile-up contributions~\cite{JETM-2018-05,JETM-2018-02}.

To resolve ambiguities when physics objects are reconstructed in close proximity, an overlap removal procedure
is applied. The two leading photons are always retained, while jets within $\Delta R = 0.4$ of a photon are removed.

The primary vertex is selected using a neural-network algorithm~\cite{HIGG-2013-08} trained to distinguish the
hard-scatter vertex of an $H\rightarrow\gamma\gamma$ event from pile-up vertices. The algorithm utilises
information from all reconstructed vertices and the kinematic properties of the diphoton system. This provides
a vertex selection efficiency of $92.1\%$ ($77.1\%$) for VBF Higgs boson events with (without) a converted
photon. The algorithm's performance was validated using $Z\rightarrow e^{+}e^{-}$ data from Run-3 $pp$ collisions by
treating electrons as unconverted photon candidates, showing good agreement between data and simulation. Following vertex
selection, the four-momenta and properties of all selected photons and jets in the event are recomputed relative to this vertex.


%

%
\section{Analysis strategy and event selection}\label{sec:Strategy}


Candidate events consistent with the $H(\rightarrow\gamma\gamma)$ + 2 jets final state are selected using
diphoton triggers~\cite{TRIG-2018-05,TRIG-2022-01}, which require the leading and subleading photons to have $\ET > 35\,\GeV$ and $\ET > 25\,\GeV$,
respectively. These triggers also apply online identification requirements based on calorimeter shower-shape variables, corresponding to the
\textit{Medium} operating point. Within the defined kinematic regions, the combined trigger efficiency for VBF Higgs boson production is around $99\%$.\par

To be considered for the final analysis, candidate events must contain at least two photons. The Higgs boson candidate is reconstructed from the two
highest-$\ET$ photons, requiring an invariant mass ($m_{\gamma\gamma}$) in the range of $105$-$160\,\GeV$. To suppress contributions from non-resonant
$\gamma\gamma$ + jets production, the leading and subleading photons must satisfy $\ET$/$m_{\gamma\gamma} >$ 0.35 and 0.25, respectively. Furthermore, at least
two jets with $\pT > 30\,\GeV$ are required. To enrich the sample with events having a VBF-like topology, the two leading jets must have a pseudorapidity gap $|\Delta\eta_{jj}| > 2.0$
and satisfy a Zeppenfeld variable~\cite{ZeppenfeldO} requirement of $\eta^\mathrm{Zepp} = \left|\eta_{\gamma\gamma} - \left(\eta_{j_1} + \eta_{j_2}\right)/2\right|<5$,
where $\eta_{\gamma\gamma}$ is the pseudorapidity of the diphoton system, and $\eta_{j_1}$ and $\eta_{j_2}$ are the pseudorapidities of the
leading and subleading jets in the event. These requirements are applied identically to both the CP and polarisation studies.\par

To further separate the VBF signal from the dominant ggF Higgs boson and $\gamma\gamma$ + jets backgrounds,
a multi-class neural network (NN) is employed. The network is trained using simulated MC samples for the VBF signal (assuming
$c_{H\widetilde{W}} = 0.0$) and the ggF background, while a data-driven approach, as described below, is used to model the continuum
background.
The network processes 16 different input features describing the VBF topology, the Higgs boson kinematics,
and the combined Higgs boson + jets system. The list of input features includes:
\begin{itemize}
\item \textbf{VBF topology} observables characterising the VBF tagged jets, including the transverse momentum and pseudorapidity of the two leading jets, the invariant dijet mass
($m_{jj}$), the pseudorapidity gap between the two jets ($\Delta\eta_{jj}$), the transverse momentum of the dijet system ($p_{\mathrm{T},jj}$), the multiplicity of
jets with $|\eta| < 2.5$ ($N^\mathrm{central\,jets}$), and the minimum angular separation between any pair of photons and jets ($\min\Delta R_{\gamma j}$).
\item \textbf{Higgs boson kinematics} observables describing the diphoton system, including the angular separation between the two photons ($\Delta R_{\gamma\gamma}$), the
diphoton transverse momentum normalised to the invariant mass ($p_{\mathrm{T},\gamma\gamma}/m_{\gamma\gamma}$), and the diphoton transverse momentum projected onto the diphoton
thrust axis normalised to the invariant mass ($p_{\mathrm{T}t,\gamma\gamma}/m_{\gamma\gamma}$).
\item \textbf{Higgs boson + jets system} observables correlating the Higgs boson and the dijet system, including the azimuthal angular difference between the two photons and the
two jets ($\Delta\Phi_{\gamma\gamma,jj}$),  the Zeppenfeld observable ($\eta^\mathrm{Zepp}$), the Collins–Soper angle of the diphoton system in the $\gamma\gamma jj$ rest frame
($\cos\Theta^{*}(\gamma\gamma,jj)$), and the transverse momentum of the Higgs boson + dijets system ($p_{\mathrm{T},\gamma\gamma jj}$).
\end{itemize}
The most important input features in the training are the $m_{jj}$, $\Delta\eta_{jj}$, $\min\Delta R_{\gamma j}$, $\Delta R_{\gamma\gamma}$, and
$p_{\mathrm{T},\gamma\gamma}/m_{\gamma\gamma}$ observables.

The classifier, implemented in PyTorch~\cite{PyTorch}, follows a sequential architecture consisting
of three hidden layers with 128, 64, and 32 nodes, respectively. Each hidden layer applies a linear transformation to the outputs of the preceding layer,
followed by layer normalisation and a ReLU activation function. The architecture concludes with a final linear layer containing three output nodes (logits)
corresponding to the VBF signal, the ggF Higgs boson background, and the non-resonant $\gamma\gamma$ + jets background. During training, the multi-class
cross-entropy loss is applied directly to the logits, while for inference the logits are transformed using a softmax function to predict the probability
of an event belonging to one of the three target processes. A dropout rate of 0.1 is applied after the second and third hidden layers to mitigate overtraining.
To prevent the classifier from sculpting the $m_{\gamma\gamma}$ spectrum, the training minimises a combined loss function: a cross-entropy term for classification
and an auxiliary distance correlation term to reduce the correlation of the network response from the $m_{\gamma\gamma}$ observable~\cite{DISCO}. The model is
optimised using the Adam algorithm~\cite{ADAM} with a learning rate of $1\times10^{-5}$ and an early stopping criterion that terminates training if the validation
loss fails to improve by at least the required tolerance of $5\times10^{-5}$ over five epochs. \par

A significant challenge in training the NN is the accurate representation of non-resonant $\gamma\gamma$ and $\gamma$ + jets backgrounds, as standard Monte Carlo
simulations often provide an insufficient description of these components in topologies typical of VBF production. To address this, a newly developed data-driven
approach based on a flow-matching method~\cite{CNF,FLOW} is used to generate the training sample for these processes. This approach uses sideband data ($m_{\gamma\gamma} \leq 120\,\GeV$ or
$m_{\gamma\gamma} > 130\,\GeV$) as input to an algorithm that learns the probability density distributions and high-dimensional correlations of all input
features to the NN training, allowing for the generation of a high-statistics, data-derived continuum background sample approximately 20 times larger than
the data sideband sample. The flow-matching model is based on a
Transformer encoder architecture with an embedded timestep, paired with a multi-layer perceptron (MLP) decoder that maps the encoded representations back to the
feature space. Training is conducted over 300 steps using the AdamW optimiser~\cite{loshchilov2019decoupledweightdecayregularization} with a learning rate
of $1\times10^{-4}$. To generate the final sample, the model is sampled in reverse steps using a Dormand–Prince solver~\cite{DorPri} with an adaptive step
size over 40 steps. This procedure results in a robust, high-statistics representation of the non-resonant background, which is essential for the subsequent
NN training and optimisation. Although trained exclusively with sideband data, studies on simulated events show that the generated sample
provides an accurate representation of the non-resonant backgrounds in the signal peak region ($120\,\GeV < m_{\gamma\gamma} \leq 130\,\GeV$).
Moreover, the underlying flow-matching model reproduces the correlation coefficients between all 16 input features to within $1\%$ of those observed in data.

To protect against potential biases, a two-fold cross-validation scheme is employed when training the NN\@. Events are randomly divided into two equal-sized
subsamples, $A$ and $B$, with independent networks trained on each. The network trained on $A$ is used to evaluate $B$, and vice-versa. Half of the data are
analysed with the NNs trained on sample $A$, and the other half with the NNs trained on sample $B$. Finally, the output distributions from both
NNs are merged for both simulated and collision data. To validate this procedure, the performance of the two independent NNs was compared on both simulation
and data, and found to be consistent within statistical uncertainties.

The predictions of the three output nodes are combined into a single discriminant:

\begin{equation}
D_{NN} = \log\left(\frac{p_\textrm{VBF}}{\left(f_\textrm{ggF}\cdot p_{ggF} + \left(1 - f_\textrm{ggF}\right)\cdot p_\textrm{Continuum}\right)}\right)
\end{equation}

where $f_\textrm{ggF}$ is the fraction of expected ggF Higgs boson events relative to the total background, which is determined to be $f_\textrm{ggF} = 0.022$.
Furthermore, the $p_{i}$ terms represent the probabilities for the various signal and background hypotheses as predicted by the NN.

The $D_{NN}$-score distribution is used to define three analysis regions with varying signal purities.
The boundaries of these regions are chosen to maximise the overall sensitivity to the Wilson coefficient $\cHWtil$. This is achieved
by iteratively moving the boundaries and performing the maximum-likelihood fits that are described in Section~\ref{sec:result}.
The configuration that yields the most stringent expected constraints on $\cHWtil$ is finally selected. The resulting intervals are
found to be equally appropriate for the polarisation study. Following this optimisation procedure, events are placed into three categories
labelled as ``tight'', ``medium'', and ``loose'' based on their $D_{NN}$-score values. The tight category, which targets the highest signal purity, is
defined by $D_{NN} \geq 3.7$, while the medium and loose categories encompass the intervals $2.1 \leq D_{NN} < 3.7$ and $-0.5 \leq D_{NN} < 2.1$, respectively.
Events with a $D_{NN}$ score below $-0.5$ are discarded, as the addition of a further low-purity category was found to provide no significant improvement to the
overall sensitivity of the two analyses. \par

Each of the three signal categories is further subdivided into eight sub-categories defined by disjoint intervals of either the $\ensuremath{\mathcal{OO}}$ or
$\Delta\Phi_{jj}$ observable, resulting in a total of 24 non-overlapping regions used per analysis. The interval boundaries for the $\ensuremath{\mathcal{OO}}$
observable are defined as $[-\infty, -0.70, -0.34, -0.15, 0.00, 0.15, 0.34, 0.70, \infty]$, while the $\Delta\Phi_{jj}$ observable is divided into eight
equidistant intervals in the range of $[0,\pi]$. Using the signal and background yields determined for each region (see Section~\ref{sec:Modelling}),
the $\ensuremath{\mathcal{OO}}$ and $\Delta\Phi_{jj}$ distributions are exploited to search for CP-violating effects and polarisation-dependent couplings,
respectively.

\begin{figure}[!h]
\centering
\includegraphics[width=0.585\textwidth]{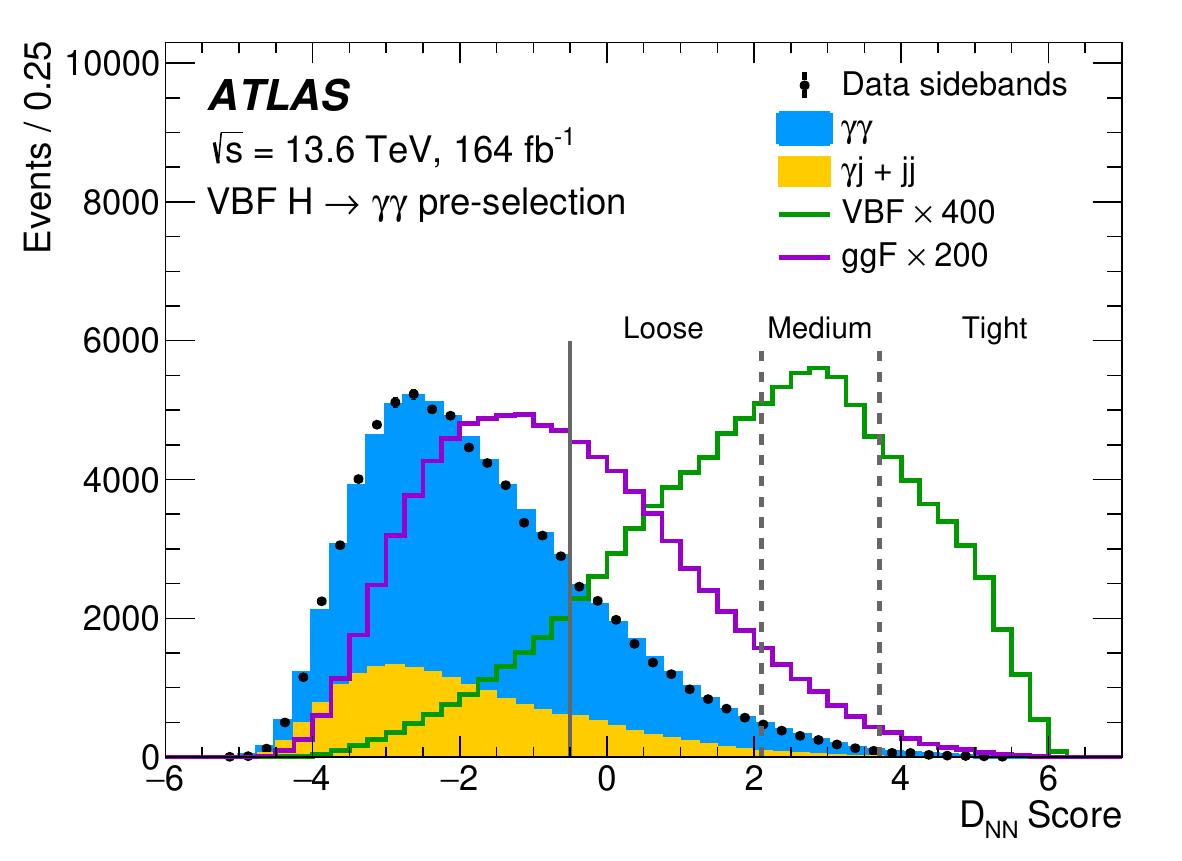}
\caption{Comparison of the NN response score ($D_{NN}$) distributions for the VBF signal, the ggF background, and the non-resonant background,
overlaid with sideband data. The non-resonant background includes simulated $\gamma\gamma$ continuum events as well as $\gamma+j$ and $jj$ processes where one or more
jets are misidentified as photons; the latter are estimated from control regions in data as detailed in Section~\ref{sec:Modelling}.
The VBF signal and ggF background are scaled by factors of 400 and 200 relative to their SM expectations, respectively.
Vertical dashed lines indicate the thresholds used to define the signal categories, while events with a $D_{NN}$ score below the lowest
threshold (solid line) are excluded from the analysis.}
\label{fig:DNN_score_distribution}
\end{figure}

The distribution of the $D_{NN}$-scores observed in the data from the $m_{\gamma\gamma}$ sidebands is presented in Figure~\ref{fig:DNN_score_distribution}.
For comparison, the figure also includes the $D_{NN}$-score templates for the non-resonant background (comprising the $\gamma\gamma$ + jets
and the $\gamma$ + jets processes obtained as detailed in Section~\ref{sec:BackgroundModelling}), the resonant ggF Higgs boson background, and the VBF signal. The boundaries of the three signal regions
are indicated by vertical dashed lines. As illustrated, the loose category is only moderately populated by the VBF signal. However, it
contains a significant fraction of events from ggF Higgs boson production. This high concentration of resonant background in the
lower-score region provides an important handle to constrain the normalisation of the ggF Higgs boson background directly from the data,
thereby reducing the systematic uncertainties associated with its contribution in the more signal-pure categories.

%

%
\section{Signal and background modelling}\label{sec:Modelling}

%

The yields for the signal, as well as the resonant and non-resonant backgrounds, are determined via a simultaneous unbinned
maximum-likelihood fit to the individual diphoton invariant mass ($m_{\gamma\gamma}$) spectra of all analysis regions. These
regions are defined by dividing the signal categories into discrete intervals of the $\ensuremath{\mathcal{OO}}$ or $\Delta\Phi_{jj}$ observables,
as described in Section~\ref{sec:Strategy}. Within each interval, the shape of the $m_{\gamma\gamma}$ distribution is modelled independently in the range
of \SIrange{105}{160}{\GeV} using analytic functional forms to describe the signal and background components.

\subsection{Signal modelling}\label{sec:SignalModelling}

The $m_{\gamma\gamma}$ distributions of the VBF signal and the resonant backgrounds from other Higgs boson production modes are modelled in each category with a
double-sided Crystal Ball (DSCB) function~\cite{Oreglia:1980cs,HIGG-2014-04}, following the approach in Ref.~\cite{HIGG-2016-21}. The function consists of a Gaussian
core describing the central peak, supplemented by two independent power-law tails at lower and higher $m_{\gamma\gamma}$ values. The total signal model includes six
shape parameters and one overall normalisation factor. Among the shape parameters, the mean and width of the Gaussian core represent the signal peak position and the
experimental resolution, respectively, while the remaining four parameters describe the tail behaviour. In each category, these six parameters are determined via fits to
simulated signal samples, comprising the various Higgs boson production modes weighted by their SM cross-sections, and are later either fixed (for the tail
parameters) or constrained within uncertainties originating from the photon energy calibration. Finally, the overall normalisation factor is determined from
the fit to data. Systematic uncertainties associated with the signal modelling are detailed in Section~\ref{sec:Sytematics}.

\subsection{Background modelling}\label{sec:BackgroundModelling}

The non-resonant background in the selected diphoton sample primarily consists of continuum $\gamma\gamma$ production, along with the $\gamma j$ and $jj$
processes where one or more jets ($j$) are misidentified as photons. For each signal category, subdivided into intervals of $\ensuremath{\mathcal{OO}}$ or $\Delta\Phi_{jj}$,
the relative contribution of each background component is measured in data using a double two-dimensional sideband method~\cite{STDM-2011-05}. This technique
uses 15 control regions in data in which either one photon (for the $\gamma j$ process) or both photons (for the $jj$ process) fail to satisfy the
identification or isolation criteria. The continuum background is on average found to be composed of approximately $79\%$ $\gamma\gamma$ events,
$19\%$ $\gamma j$, and $2\%$ $jj$ events. The $\gamma\gamma$ fraction changes smoothly across the analysis regions and ranges from $66\%$ to $97\%$. \par

The $m_{\gamma\gamma}$ shape for the continuum $\gamma\gamma$ background is derived from simulated background samples after applying the
full event selection described in Section~\ref{sec:Strategy}. In contrast, the shapes for the $\gamma j$ and $jj$ components are obtained from data control
samples defined by inverting the identification requirements for one or both photons, while keeping all other selection criteria. The ratio of the
$m_{\gamma\gamma}$ distributions of these data-driven components to those of the simulated $\gamma\gamma$ sample is well-described by a
second-order polynomial. A fit to this ratio is used to derive a $m_{\gamma\gamma}$-dependent weight, which is applied to the simulated
$\gamma\gamma$ sample to produce a final set of background templates that incorporate the contributions and kinematics of the $\gamma j$ and $jj$ processes.
These templates, defined over the range $105 \leq m_{\gamma\gamma} < 160\,\GeV$ in 55 uniform-width bins, are used solely to evaluate the spurious signal and
select the background functional forms as described below. The final analysis results are then obtained by fitting the chosen analytic functions directly to the data.

Three classes of analytic functions are evaluated as candidates for modelling the non-resonant background in each category. They include exponential functions
of first- to third-order polynomials, Bernstein polynomials of third to fifth order~\cite{BernStein}, and first-order power-law functions. The background function for each
$\ensuremath{\mathcal{OO}}$ or $\Delta\Phi_{jj}$ region is selected independently based on the magnitude of the ``spurious signal''~\cite{HIGG-2012-27} and the overall fit quality.
The spurious signal, defined as the systematic bias in the fitted signal yield resulting from the functional form choice, is estimated by performing signal-plus-background
fits to the background templates. This procedure is repeated by scanning the Higgs boson mass distribution in the range of \SIrange{121}{129}{\GeV}
in steps of \SI{0.5}{\GeV}, with the absolute maximum of the fitted signal yield across the scan taken as the spurious signal value. \par

To be considered, a function must satisfy two criteria: a $\chi^2$ p-value greater than $0.1\%$ and a spurious signal yield less than
$20\%$ of the statistical uncertainty of the fitted signal yield or $10\%$ of the expected signal yield. If multiple functions meet these
requirements, the one with the fewest degrees of freedom is chosen. In cases where no function initially satisfies the spurious signal
criteria, the $\chi^2$ requirement is removed. For both the CP and polarisation studies, these criteria were not met in two out of the
24 analysis regions due to large statistical fluctuations in the background templates. In these instances, the candidate functions were
instead fitted directly to the sideband data to confirm they provided a sufficiently accurate description of the $m_{\gamma\gamma}$ shape.
The final spurious signal value is incorporated as a systematic uncertainty in the signal yield to account for potential biases in the
background modelling. Furthermore, uncertainties in the $\gamma j$ and $jj$ fractions are found to have a negligible impact on the outcome
of the spurious-signal test and are therefore neglected in the final fit.


%

%
\section{Systematic uncertainties}\label{sec:Sytematics}


Systematic uncertainties considered in the two analyses can be grouped into two main areas: uncertainties in
the modelling of the $m_{\gamma\gamma}$ distribution for the signal and background processes, and uncertainties in the
predicted VBF signal and ggF background yields in each category arising from experimental or theory sources. These systematic
uncertainties are incorporated into the likelihood model as nuisance parameters (NPs)~\cite{Conway:2011in}, as
detailed in Section~\ref{sec:result}. Further information about the relevant uncertainties is provided below.

\subsection{Experimental systematic uncertainties}

Most experimental uncertainties primarily affect the expected yields of the VBF signal and the ggF background.
The photon identification and isolation efficiencies are corrected to match those observed in data using dedicated measurements~\cite{EGAM-2021-01}.
These correction factors, along with their corresponding uncertainties, are determined as a function of the photon $\ET$ and $|\eta|$ using control
samples of prompt photons from $\gamma$ + jets events, radiative $Z\rightarrow \ell^+ \ell^-\gamma$ decays, and electrons from $Z\rightarrow e^+ e^-$
decays. In the latter case, electrons are used as a proxy for photons due to their similar electromagnetic shower signatures,
with Smirnov transformations applied to account for residual differences between their respective shower-shape distributions.
The impact on the event yields from uncertainties in the photon identification and isolation efficiencies range from $3.6\%$ to $5.7\%$ and from $1.6\%$ to $2.0\%$, respectively, for both
the VBF signal and the ggF background. The uncertainty in the efficiency of the diphoton trigger is estimated by using radiative $Z$ boson decays and events
from pre-scaled lower-threshold triggers~\cite{TRIG-2018-05}. The resulting uncertainties are typically around $0.2\%$.

The photon energy scale and resolution are corrected to account for energy loss in upstream and inactive material, leakage effects, and imperfect modelling of
electromagnetic showers and electronics response. These corrections, and the relevant uncertainties, are derived from control samples of electrons from $Z$
boson decays and of photons from radiative $Z\rightarrow \ell^+ \ell^-\gamma$ decays~\cite{EGAM-2021-02}. The resulting uncertainties in the predicted yields of
the VBF signal and the ggF background range from $0.2\%$ to $0.5\%$.

Uncertainties in the jet energy scale (JES) and resolution (JER) are estimated by using control samples where jets recoil
against well-calibrated reference objects, such as a photon or a $Z$ boson. These uncertainties incorporate contributions from
\textit{in situ} calibration measurements, the dependence on pile-up activity, and the jet flavour composition~\cite{JETM-2018-05}.
To account for an observed mismodelling of jets at large pseudorapidity ($|\eta| > 3.2$) resulting from an energy shift measured in the forward
calorimeters, an additional systematic uncertainty is introduced. Furthermore, JES and JER uncertainty terms are included to account for the
non-closure between the G4 and AF3 simulations. The performance of the jet vertex tagging
algorithm, employed to suppress contributions from pile-up jets, is corrected to match the efficiencies observed in
data~\cite{PERF-2014-03}. The associated corrections factors and their systematic uncertainties are fully
propagated through both studies. Summing these individual contributions, the total uncertainty in the VBF signal yields ranges from $4\%$ to $11\%$, while the
ggF background uncertainty ranges from $7\%$ to $25\%$ across all analysis regions.

Uncertainties in the modelling of pile-up activity in the simulation are accounted for by varying the value of the visible inelastic cross-section used
for reweighting the simulation to match the data~\cite{STDM-2015-05}. The resulting uncertainties in the VBF signal and the ggF background yields range from
$0.1\%$ to $1.4\%$.

Dedicated uncertainties are evaluated to account for residual discrepancies between the G4 and AF3 simulations. These uncertainties
are determined by comparing the predicted event yields of the VBF signal and the ggF background using samples where the calorimeter responses are simulated
with either AF3 or G4. The comparisons are performed independently for each signal category and within each $\ensuremath{\mathcal{OO}}$ and
$\Delta\Phi_{jj}$ interval. The resulting yield differences are assigned as additional systematic uncertainties, which are typically on the order of $5\%$ for
both processes.

Finally, a global luminosity uncertainty of $1.9\%$ is applied to the normalisation of the simulated signal
and background samples. This uncertainty is determined following the methodology discussed in
Ref.~\cite{ATL-DAPR-PUB-2025-001}, using the LUCID-2 detector~\cite{LUCID2} for the primary luminosity measurements.

Beyond their impact on the expected event yields, the uncertainties in the energy scale and energy resolution of the photon candidates also
affect the shape of the $m_{\gamma\gamma}$ distribution
for both the VBF signal and the Higgs boson background processes. Most notably, these include uncertainties in the energy scale and energy resolution of the
photon candidates. The photon energy scale uncertainties are propagated as shifts to the peak position of the DSCB function. The resulting impact is typically
less than $0.3\%$ relative to the nominal peak position, depending on the specific signal category and the $\ensuremath{\mathcal{OO}}$ or $\Delta\Phi_{jj}$ interval.
Similarly, the photon energy resolution uncertainties are propagated to the Gaussian width of the DSCB shape. This variation has a relative impact ranging from
$6\%$ to $14\%$, depending on the event category and the corresponding $\ensuremath{\mathcal{OO}}$ or $\Delta\Phi_{jj}$ interval.

The modelling of the non-resonant $\gamma\gamma$ + jets background is subject to the choice of the analytic
functional form. The resulting uncertainties in the background estimate for each interval of the $\ensuremath{\mathcal{OO}}$ and \DPhijj\ distributions
are quantified via the spurious signal yield, as detailed in Section~\ref{sec:Modelling}. For each of the tested background functions,
the spurious signal is smaller than $30\%$ of the statistical uncertainty of the expected signal yield.

\subsection{Theoretical systematic uncertainties}

Uncertainties in the theoretical modelling of the simulated VBF signal and the ggF background processes are taken into account. These modelling
uncertainties arise from missing higher-order terms in the perturbative QCD calculations, the modelling of parton showers and the PDF set,
and uncertainties in the value of the strong coupling constant $\alpha_{S}$.

Uncertainties in the modelling of the PDF sets are evaluated following the PDF4LHC recommendations~\cite{Butterworth:2015oua}, while the impact of the $\alpha_{S}$
uncertainty is assessed by varying the nominal value ($\alpha_{S}=0.118$) by $\pm 0.001$. These effects result in yield variations of approximately $2\%$ and $0.5\%$,
respectively.

Furthermore, uncertainties due to missing higher-order terms in the perturbative expansion are evaluated by varying the renormalisation and
factorisation scales, $\mu_{R}$ and $\mu_{F}$. For the VBF signal process, these uncertainties are estimated by individually varying
$\mu_{R}$ and $\mu_{F}$ by a factor of $2$ or $0.5$ relative to their nominal value. Six combinations are considered: $(\mu_{R}, \mu_{F})$ $=$ $(0.5, 0.5)$,
$(0.5, 1.0)$, $(1.0, 0.5)$, $(1.0, 2.0)$, $(2.0, 1.0)$, and $(2.0, 2.0)$ times their nominal value. The final uncertainty is defined as the envelope of these
variations, taking the maximal upward and downward deviations from the nominal predictions. These variations result in an uncertainty of up to $10\%$ on the
expected signal yields. For the ggF background process, a simple variation of the renormalisation and factorisation scales is insufficient, as it tends to
underestimate the actual uncertainties. Instead, the procedure detailed in Ref.~\cite{ATL-PHYS-PUB-2023-031} is followed. This approach provides 18 individual
nuisance parameters covering effects from jet migration, the Higgs boson \pt shape, VBF topology selection, and top-quark-mass dependence.
These uncertainties are typically less than $15\%$ of the expected ggF background yields. \par

The modelling of the parton shower, underlying event, and hadronisation is assessed separately for the VBF signal and ggF background processes.
This is achieved by replacing \Pythia{8.310} with \Herwig{7.2.3} with the H7UE set of tuned parameters~\cite{Bellm:2019zci} for the event simulation.
The uncertainties estimated from the differences between the resulting yield predictions typically do not
exceed $10\%$ for the VBF signal and $15\%$ for the ggF background process.

Since the other Higgs boson production modes ($ZH$, $WH$, $b\bar{b}H$, $t\bar{t}H$, $tWH$, and $tHjb$) contribute only minimally to the signal
regions probed in the two analyses, only uncertainties in their cross sections are taken into account. These uncertainties include effects from
varying the renormalisation and factorisation scales, the PDF set, and $\alpha_\mathrm{S}$. The exact values are taken from
Ref.~\cite{deFlorian:2016spz}.


%

%
\section{Results}\label{sec:result}

%

An unbinned maximum-likelihood fit is used for the statistical interpretation of the results from both the CP- and polarisation studies.
Fits are performed independently for each parameter of interest (POI): $c_{H\widetilde{W}}$, $a_\mathrm{L}$, and $a_\mathrm{T}$.
In each case, a simultaneous fit is conducted on the $m_{\gamma\gamma}$ distributions across the 24 analysis regions.
Systematic uncertainties are implemented as nuisance parameters (NPs) with their correlations maintained across regions. The normalisations
of the VBF signal and ggF background processes are parameterised via the normalisation scale factors, $\mu_{\mathrm{VBF}}$ and $\mu_{\mathrm{ggF}}$,
which are treated in most fit configurations as free parameters common to all regions. The normalisations of the other Higgs boson production processes are
constrained to the SM predictions within their respective theoretical uncertainties.

The expected performance is evaluated using Asimov data samples~\cite{Cowan:2010js} generated under the SM hypothesis. For fits to the Asimov data sample, the
background shape and yield are fixed based on a fit to the data sidebands, excluding the signal peak region (\SIrange{120}{130}{\GeV}).
As described in Section~\ref{sec:DataAndMC}, parameter morphing is used to interpolate between a discrete set of simulated
coupling values for $c_{H\widetilde{W}}$, $a_\mathrm{L}$, and $a_\mathrm{T}$ to model the signal across a continuous range of coupling scenarios.

To obtain the final results, the fitting procedure is applied individually to each coupling parameter hypothesis, where the background prediction is only
affected through variations of the normalisation factors and the NPs. A negative log-likelihood (NLL) curve is constructed as a function of the
relevant coupling parameters. The value at the NLL minimum is taken as the best-fit estimate of the corresponding POI, while the associated confidence
level (CL) intervals are determined from the appropriate deviation of the NLL from its minimum.

\subsection{Run-3 results} \label{sec:results_run3}

In the fit configurations for the CP studies, the normalisation scale factors $\mu_{\mathrm{VBF}}$ and $\mu_{\mathrm{ggF}}$ are treated as free-floating parameters.
By allowing these parameters to float, the analysis relies solely on the shape of the $\ensuremath{\mathcal{OO}}$ distribution, thereby reducing sensitivity to BSM CP-even
contributions that could modify the total cross-section. However, as the quadratic BSM terms can modulate the $\ensuremath{\mathcal{OO}}$ distribution in a symmetric manner,
they may still influence the measured shape. Consequently, the constraints on $c_{H\widetilde{W}}$ are derived using two distinct parameterisations: one including
only the linear terms in the morphing procedure, and another accounting for both the linear and quadratic terms. \par

For the polarisation studies, the parameters $a_\mathrm{L}$ and $a_\mathrm{T}$ are determined in two separate fit configurations. When one parameter is treated as the
POI, the other is fixed to its SM value. Two distinct types of fits are considered: a shape-only fit, where both $\mu_{\mathrm{VBF}}$ and $\mu_{\mathrm{ggF}}$
are treated as free parameters, and a shape+rate fit, where $\mu_{\mathrm{ggF}}$ is floated while $\mu_{\mathrm{VBF}}$ is fixed to unity.
The shape-only fit relies exclusively on the shape information of the $\Delta\Phi_{jj}$ distribution to distinguish between different signal hypotheses, whereas
the shape+rate fit incorporates both the distribution shape and the total event yield. In the shape+rate fit, the signal normalisation is not free to
adjust to the data but instead follows the cross section predicted for each ($a_\mathrm{L}$, $a_\mathrm{T}$) hypothesis, allowing rate information to contribute to
the constraints of the POI. The two parameters show distinct sensitivities: $a_\mathrm{L}$ significantly
modifies the total yield because the longitudinal polarisation vectors of the massive gauge bosons are proportional to the energy and can lead to large enhancements
in the total cross section, whereas $a_\mathrm{T}$ has a comparatively weaker impact on the normalisation. Consequently, $a_\mathrm{L}$ is more strongly constrained by
rate information, while the sensitivity to $a_\mathrm{T}$ is primarily driven by the shape of the $\Delta\Phi_{jj}$ distribution. \par

\begin{figure}[h!]
\centering
\subfloat[]{\includegraphics[width=0.475\textwidth]{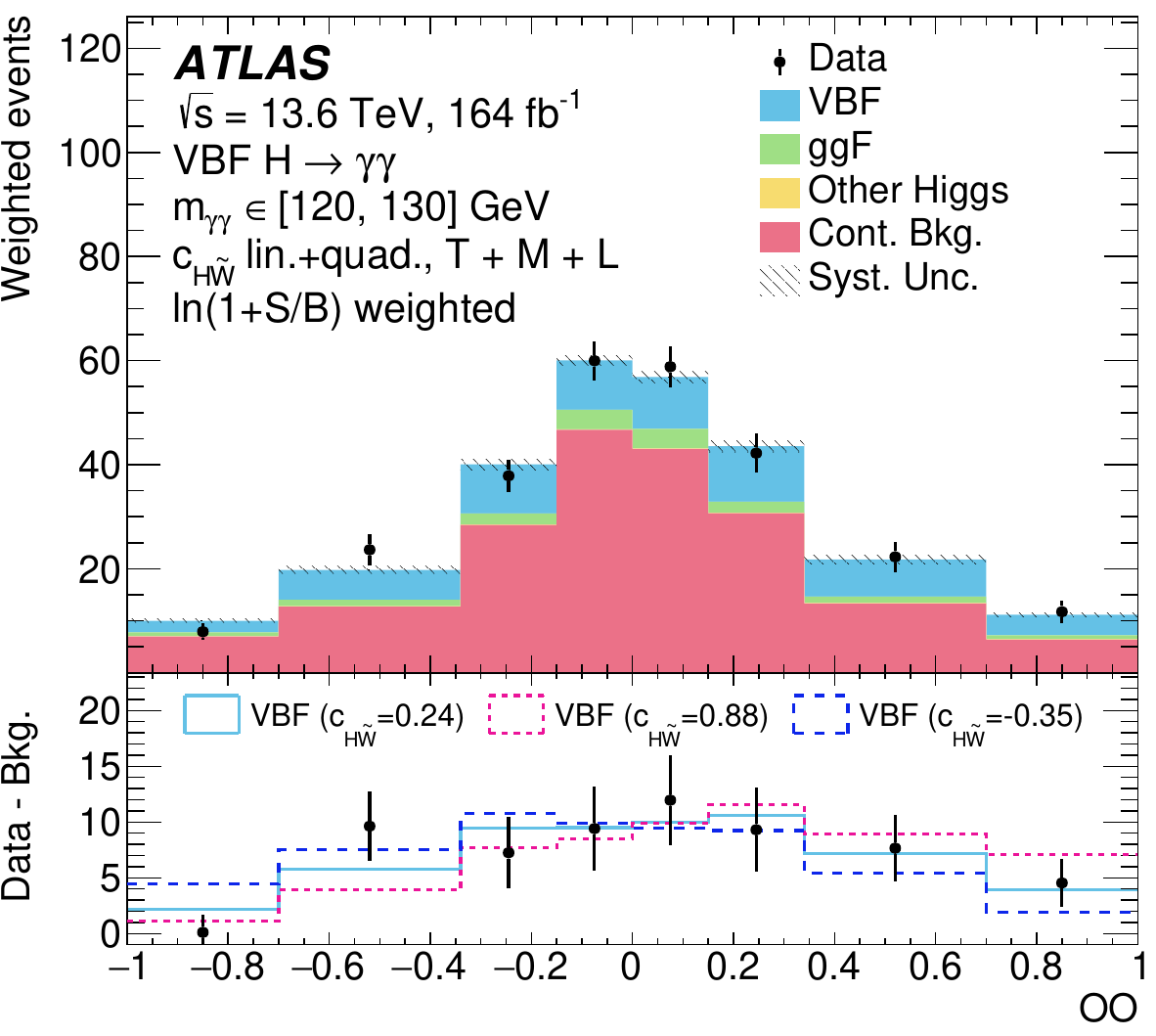}}
\subfloat[]{\includegraphics[width=0.475\textwidth]{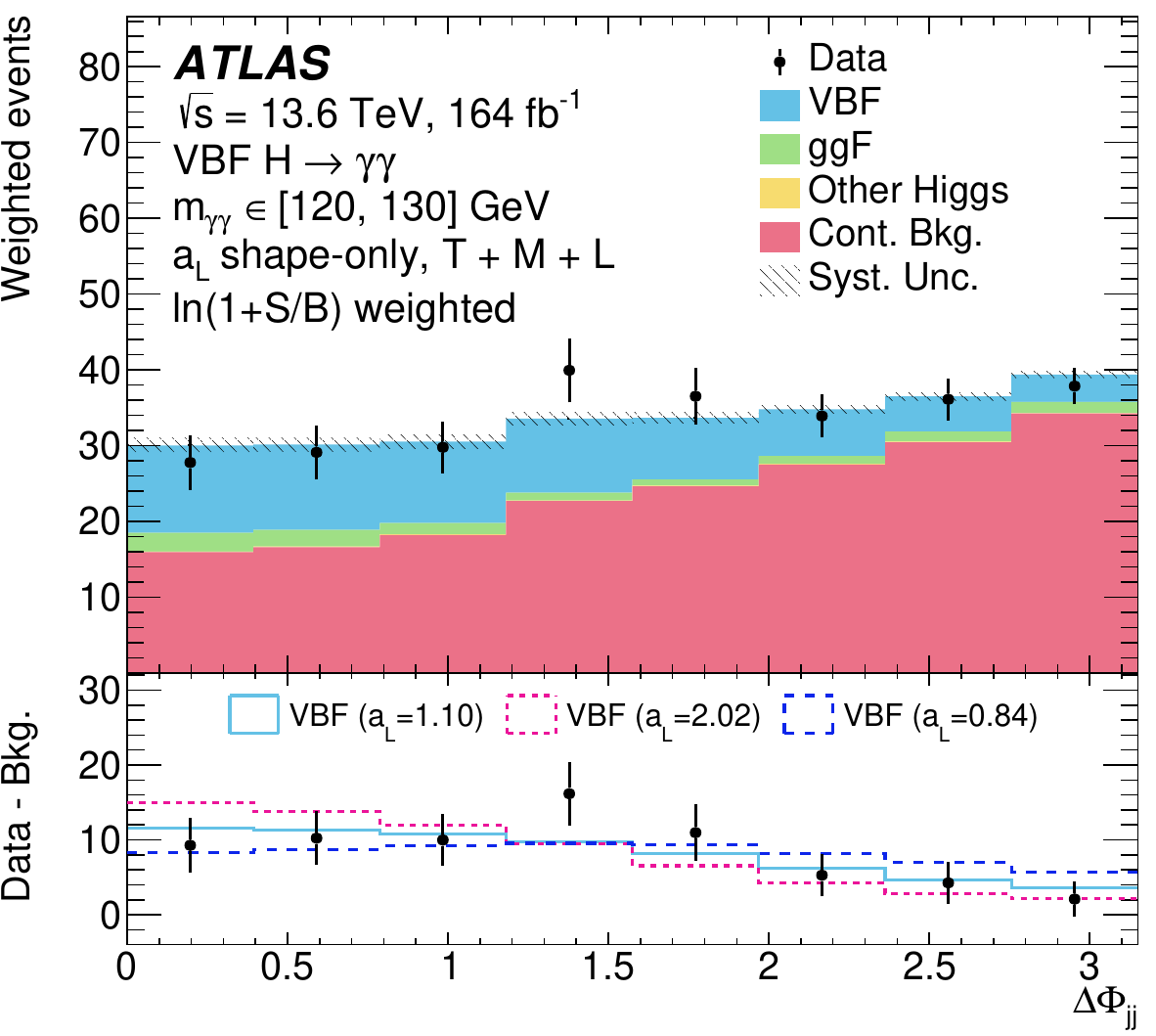}}
\caption{Post-fit distributions of (a) the optimal observable and (b) $\Delta\Phi_{jj}$ for events in the combined
tight (T), medium (M), and loose (L) analysis regions within the invariant diphoton mass window $m_{\gamma\gamma}\in [120,130]\,\GeV$.
The signal and background yields are fixed to the best-fit values from the $c_{H\widetilde{W}}$ and $a_\mathrm{L}$ interpretations.
Contributions from the three regions are summed, weighted by $\ln(1+S/B)$, where $S$ and $B$ are the best-fit signal and background
yields, respectively. The overflow and underflow are included in the outermost bins. The uncertainty band includes all systematic
uncertainties. The lower panels show the background-subtracted data compared with the best-fit VBF prediction and two BSM scenarios
corresponding to the two POI values excluded at exactly $95\%$ CL by fits using only shape information.}
\label{fig:Distribution_VBF_weighted}
\end{figure}

\begin{figure}[h!]
\centering
\includegraphics[width=0.605\textwidth]{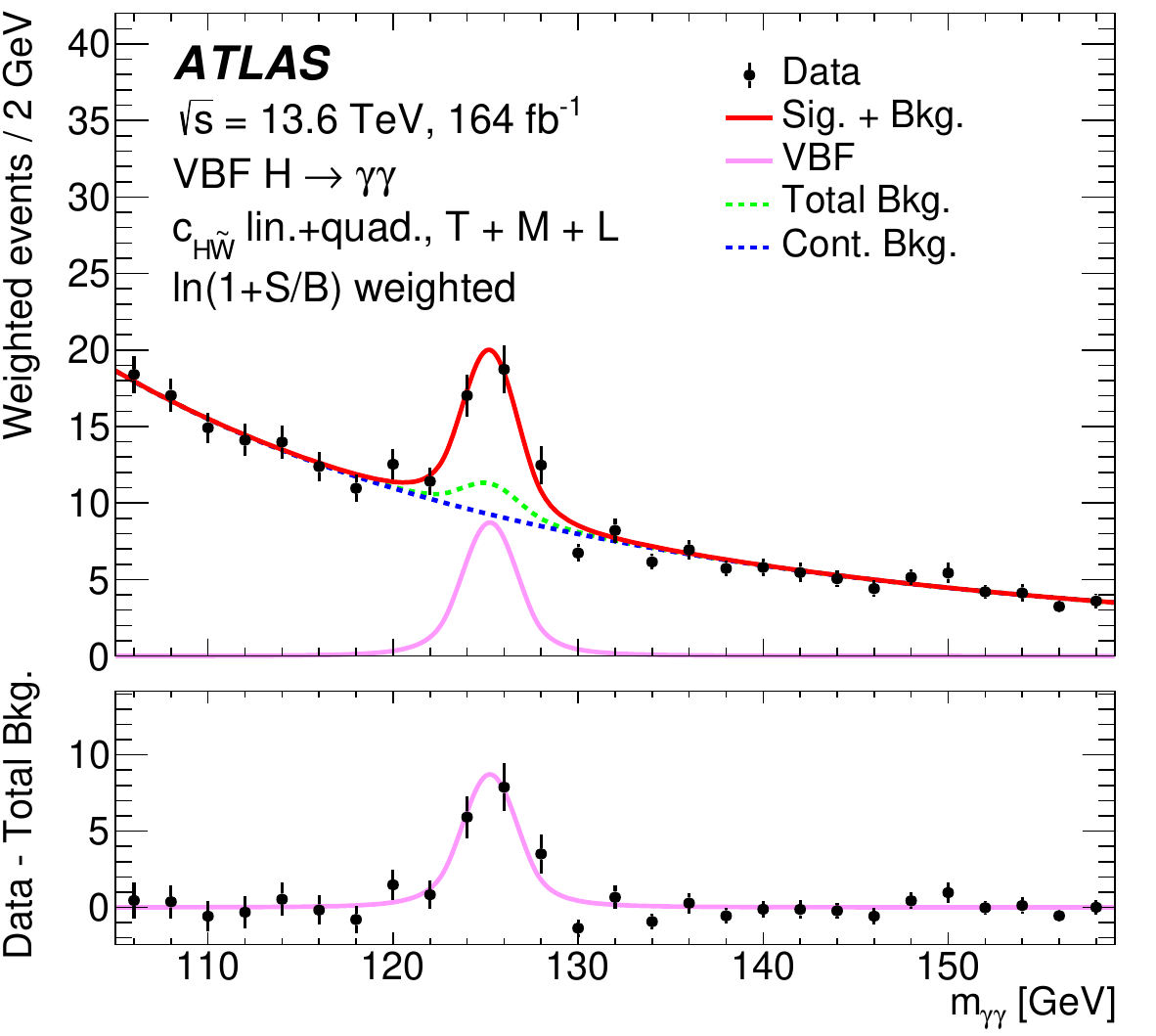}
\caption{Weighted post-fit distribution of the data events compared with the signal and background contributions for the
$m_{\gamma\gamma}$ spectrum, summed over the tight (T), medium (M), and loose (L) analysis regions and optimal observable intervals.
Events are weighted by $\ln(1+S/B)$, where $S$ and $B$ are the best-fit signal and background yields in each analysis region and optimal
observable interval. The signal and background contributions are fixed to the values obtained from the best fit at
$c_{H\widetilde{W}}=0.24$\@.}
\label{fig:Distribution_myy_CP}
\end{figure}

Figure~\ref{fig:Distribution_VBF_weighted} depicts the weighted $\ensuremath{\mathcal{OO}}$ and $\Delta\Phi_{jj}$ distributions summed
across all three signal categories. For each category, events are weighted by $\ln(1+S/B)$, where $S$ and $B$ are the corresponding
post-fit yields of signal and background events, respectively. The post-fit distribution of the diphoton invariant mass summed over all
analysis categories and the intervals of the optimal observable is shown in Figure~\ref{fig:Distribution_myy_CP}, where the events in
each category are weighted by $\ln(1+S/B)$. The signal and background contributions are fixed to the best-fit values from the
$c_{H\widetilde{W}}$ interpretations.\par

\begin{figure}[h!]
\centering
\subfloat{\includegraphics[width=0.605\textwidth]{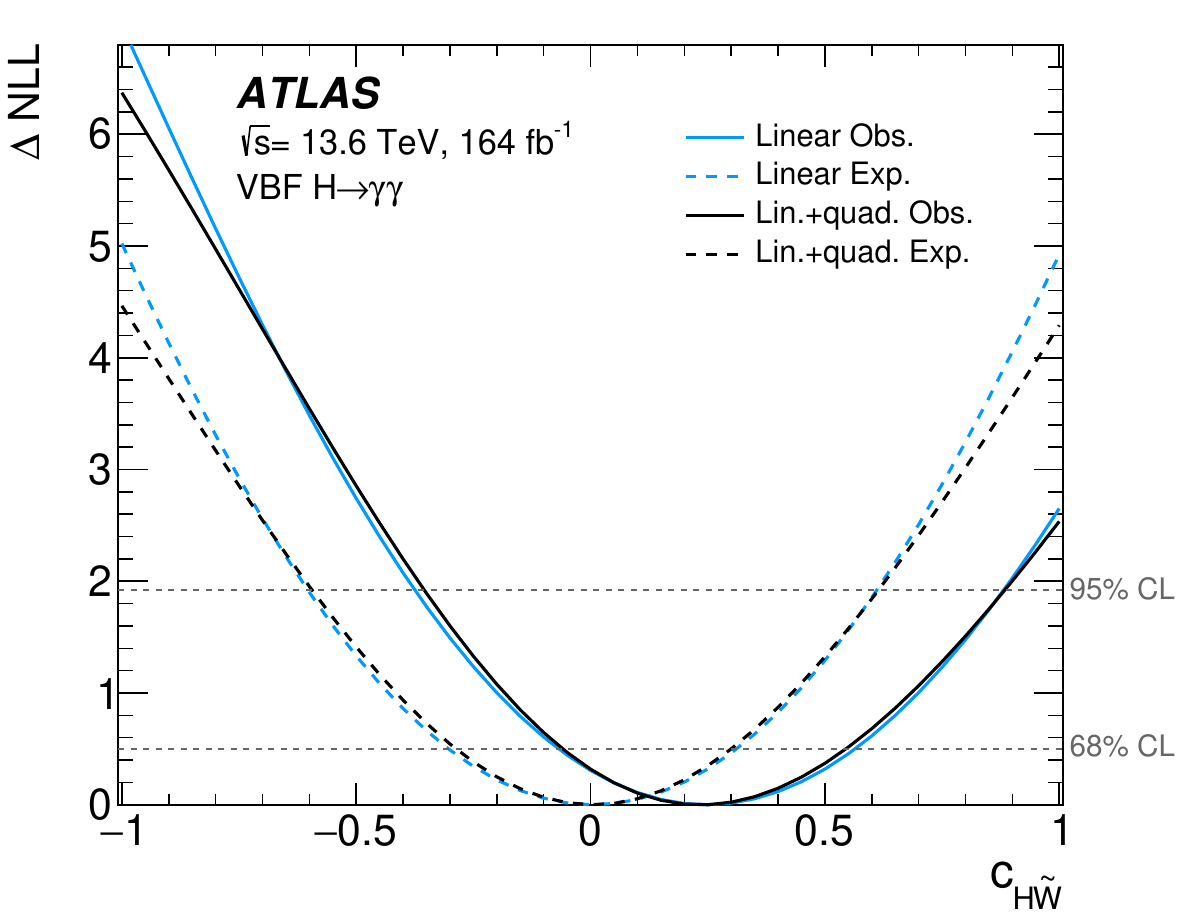}}
\caption{Expected and observed likelihood curves as a function of the Wilson coefficient $c_{H\widetilde{W}}$. Fits using parameterisations with linear-only
terms and linear+quadratic terms are shown. The dashed horizontal lines indicate the $\Delta$NLL values corresponding to the $68\%$ and $95\%$ CL intervals.
Results are obtained assuming a new physics scale of $\Lambda = 1\,\TeV$.}
\label{fig:Disctribution_NLL_cp}
\end{figure}

\begin{figure}[h!]
\centering
\includegraphics[width=0.475\textwidth]{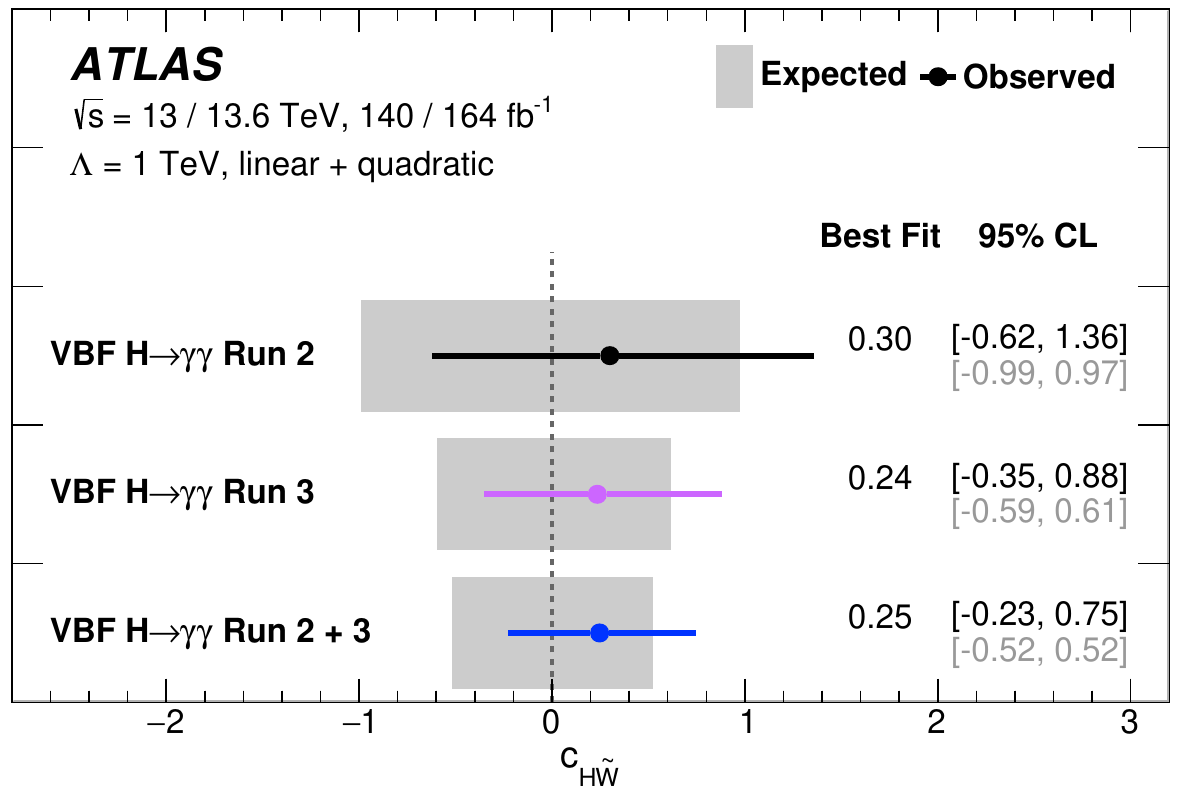}\\
\caption{The expected and observed measurements of the Wilson coefficient $c_{H\widetilde{W}}$ in the VBF $H\rightarrow \gamma\gamma$ channel.
Results are shown for integrated luminosities of 140 fb$^{-1}$ at $\sqrt{s}=13$\,\TeV~\cite{HIGG-2020-08} and 164 fb$^{-1}$ at $\sqrt{s}=13.6$\,\TeV,
and for the statistical combination of both data samples (as presented in Sec.~\ref{sec:results_combination}). The dots and $95\%$ CL
bars show the best-fit values with statistical and systematic uncertainties, and the grey bands centred at zero show the expected $95\%$ CL
intervals. The best fit values and $95\%$ CL observed (expected) limits are also explicitly stated on the right-hand side of the plot in black (grey).
Results are obtained assuming a new physics scale of $\Lambda = 1\,\TeV$.}
\label{fig:Summary_of_results_cp}
\end{figure}

The results of the likelihood scans for the CP studies are presented in Figure~\ref{fig:Disctribution_NLL_cp}. The resulting best-fit value, as well as the expected and
observed $95\%$ CL intervals, are detailed in Figure~\ref{fig:Summary_of_results_cp} for the fit using the linear+quadratic parameterisation. The corresponding
normalisation scale factors for the VBF signal and ggF background, at the best-fit estimate of $c_{H\widetilde{W}}$, are found to be
$\mu_\mathrm{VBF} = 1.01 ^{+0.24}_{-0.20}$ and $\mu_\mathrm{ggF} = 1.27 ^{+0.56}_{-0.45}$, respectively. The $68\%$ ($95\%$) CL limits with linear-only terms shift
by approximately $3.0\%$ ($8.6\%$), indicating a negligible impact of the quadratic terms on the exclusion limits.
Figure~\ref{fig:Summary_of_results_cp} also shows a comparison of the expected and observed best-fit values and $95\%$ CL intervals of the Run-2
VBF $H\rightarrow \gamma\gamma$ analysis and the statistical combination of the Run-2 and Run-3 results. \par

\begin{figure}[h!]
\centering
\subfloat[]{\includegraphics[width=0.375\textwidth]{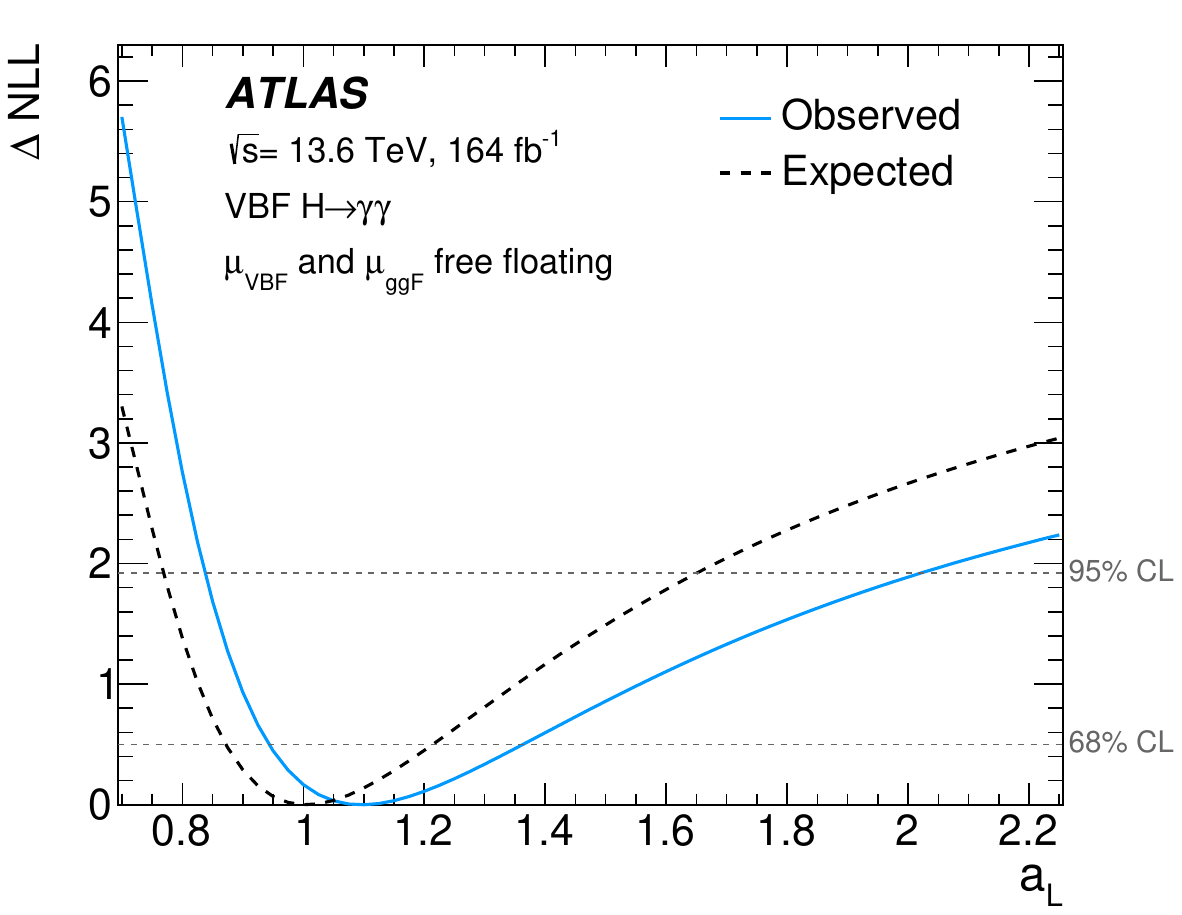}}
\subfloat[]{\includegraphics[width=0.375\textwidth]{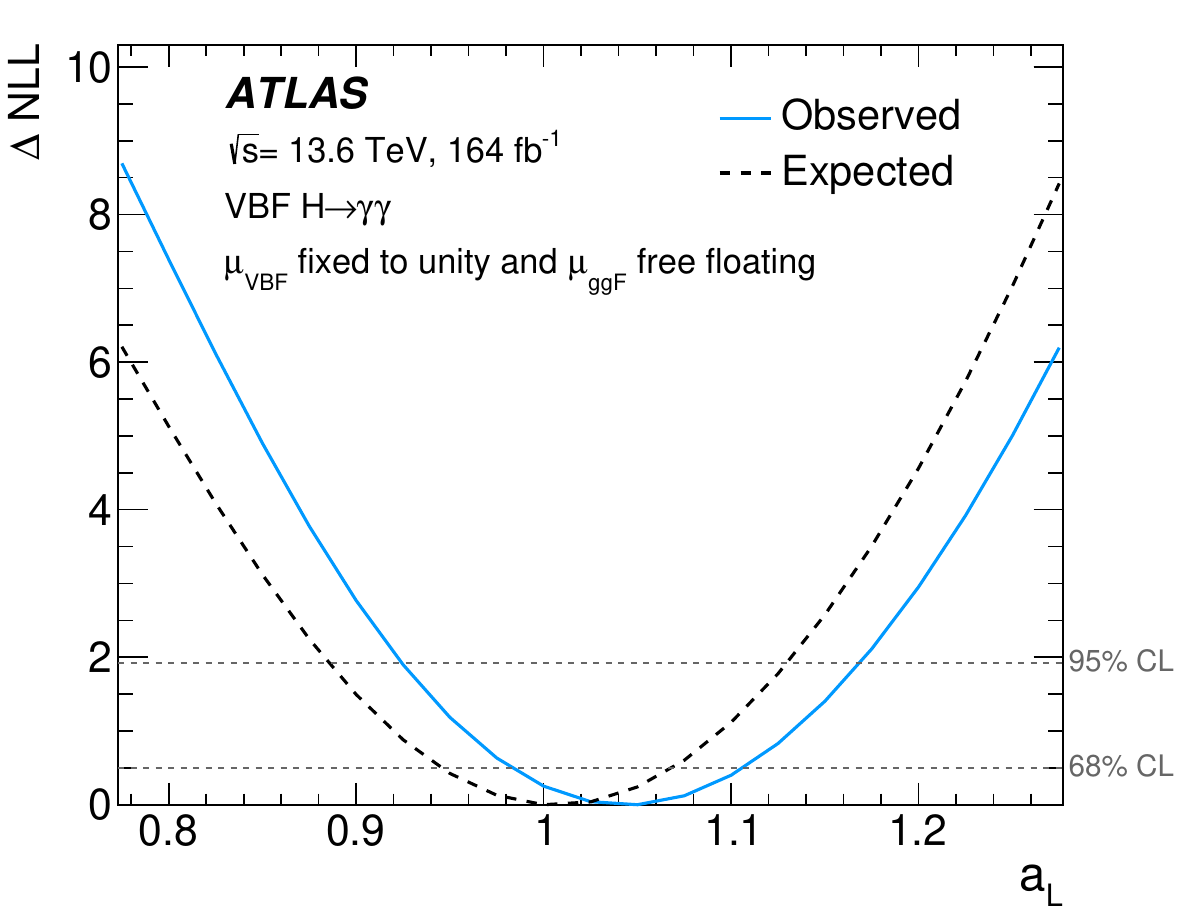}}\\
\subfloat[]{\includegraphics[width=0.375\textwidth]{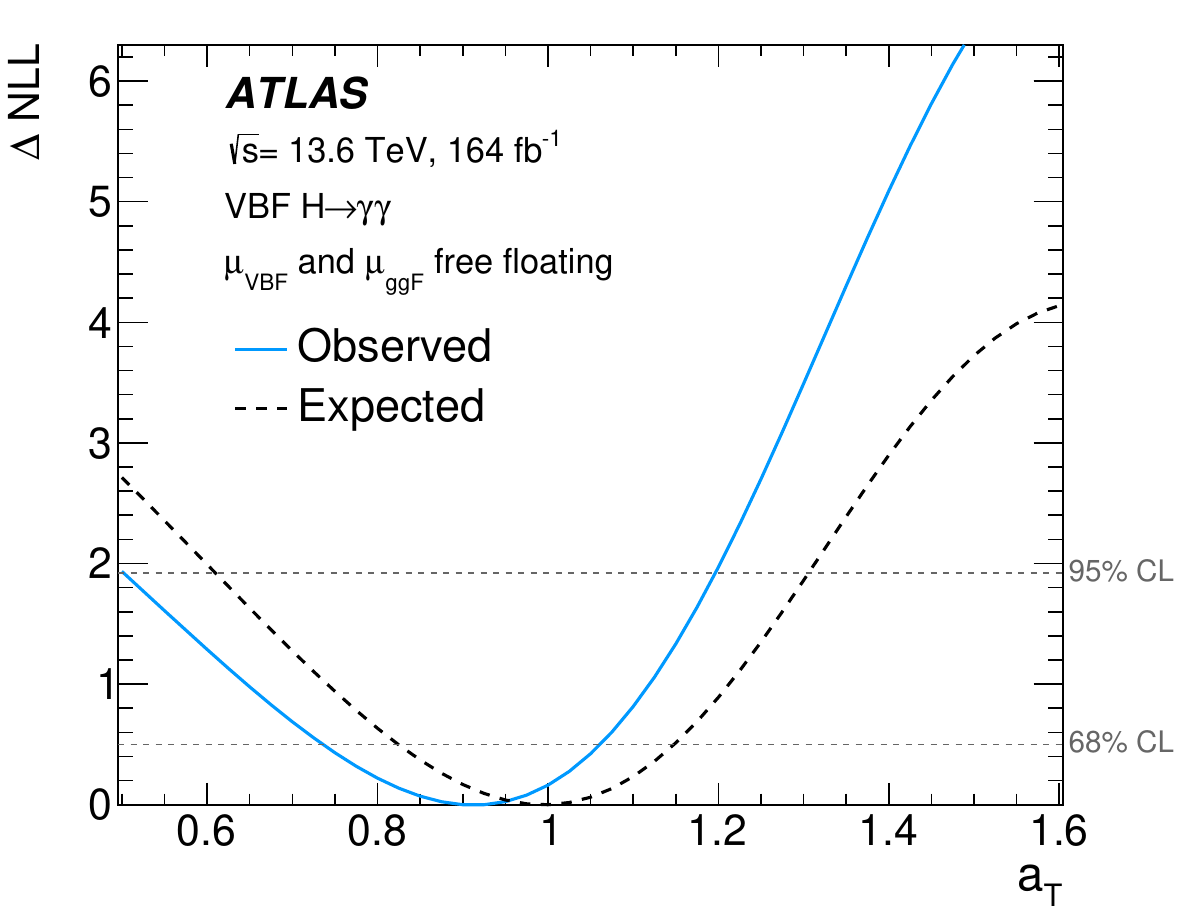}}
\subfloat[]{\includegraphics[width=0.375\textwidth]{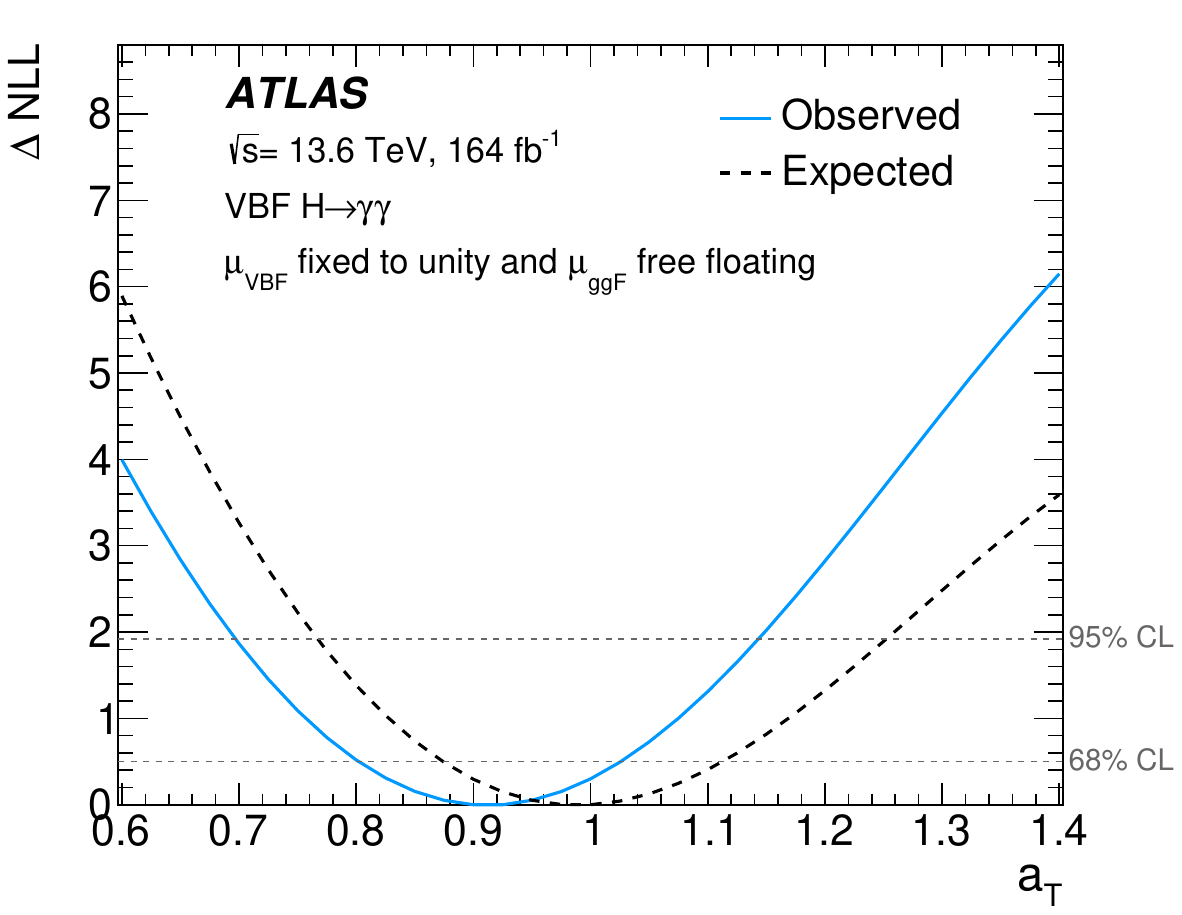}}
\caption{Expected and observed likelihood curves as a function of the longitudinally (a, b) and transversely (c,d) polarised couplings.
Fits using shape-only (a,c) and shape+rate (b, d) information are shown.  For all figures, the dashed horizontal lines indicate the
$\Delta$NLL values corresponding to the $68\%$ and $95\%$ CL intervals.}
\label{fig:Disctribution_NLL_pol}
\end{figure}

\begin{figure}[h!]
\centering
\includegraphics[width=0.575\textwidth]{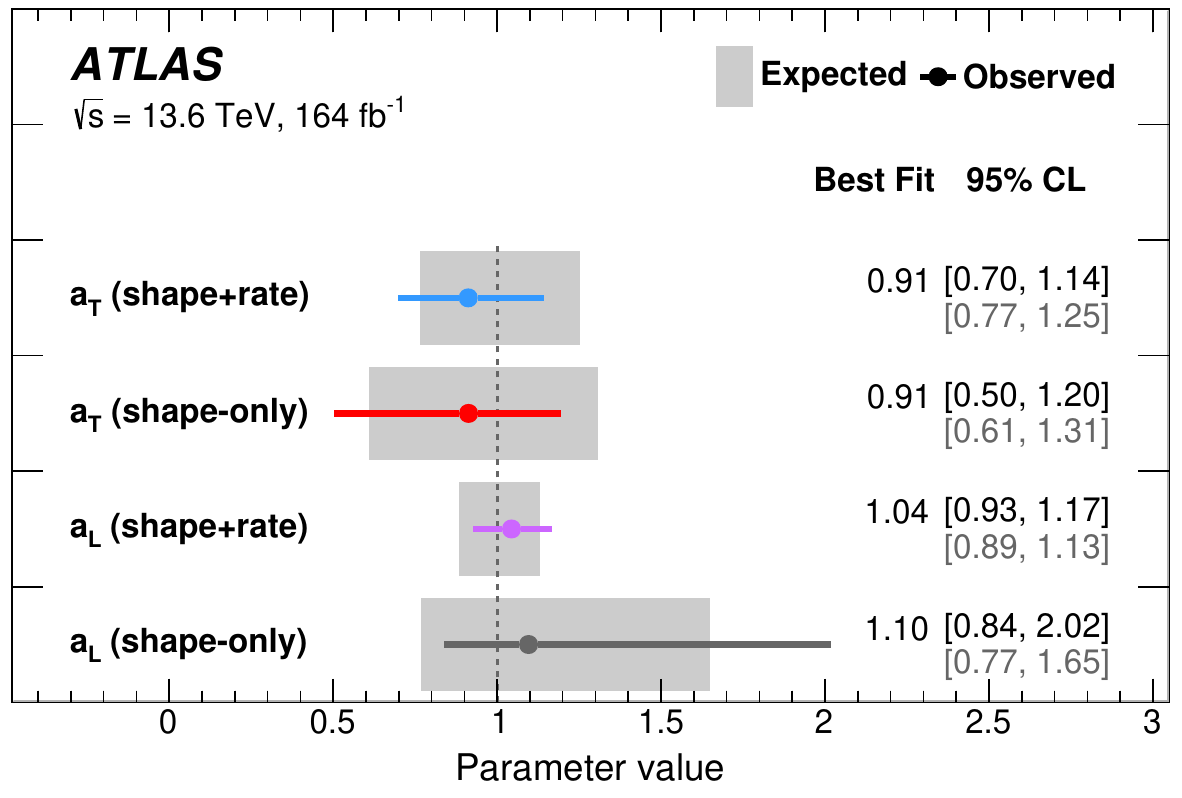}
\caption{The expected and observed measurements of the polarisation-dependent coupling-strength scale factors for an integrated luminosity of 164 fb$^{-1}$
at $\sqrt{s}=13.6$\,\TeV.  The dots and $95\%$ CL bars show the best-fit values with statistical and systematic uncertainties, and the grey bands
centred at one show the expected $95\%$ CL intervals. The best fit values and $95\%$ CL observed (expected) limits are also explicitly stated
on the right-hand side of the plot in black (grey).}
\label{fig:Summary_of_results_pol}
\end{figure}

The results of the likelihood scans for the polarisation studies are presented in Figure~\ref{fig:Disctribution_NLL_pol}. The
scans over $a_\mathrm{L}$ ($a_\mathrm{T}$) are shown in the upper (lower) panel, and the resulting best-fit values and the $95\%$ CL intervals are detailed
in Figure~\ref{fig:Summary_of_results_pol}. The corresponding normalisation factors $\mu_\mathrm{VBF}$ and $\mu_\mathrm{ggF}$ are consistent within one standard
deviation with those obtained from the CP studies.
The asymmetry of the likelihood curves from the shape+rate fits stems from the non-linear dependence of the cross-section on the coupling parameters
(see Ref.~\cite{Brehmer:2014pka}). Conversely, the asymmetry in the shape-only fits is primarily driven by the background distribution, as the
concentration of background at low $\Delta\Phi_{jj}$ values in the most sensitive signal category creates a lopsided sensitivity. For $a_\mathrm{L} < a_\mathrm{T}$,
the VBF signal shifts toward larger $\Delta\Phi_{jj}$ values where the background is low, whereas for $a_\mathrm{L} > a_\mathrm{T}$, the signal overlaps more
significantly with the background at low $\Delta\Phi{jj}$ values. \par

The results from both the CP and polarisation studies are consistent with the SM expectations.\par

A breakdown of the impacts of various uncertainty sources on each POI, evaluated at the $68\%$ CL, is summarised in
Table~\ref{tab:syst_breakdown}. For all fits, the statistical uncertainty is the dominant component. Among the leading
systematic uncertainties are those related to the modelling of the VBF signal process, the spurious signal, the energy
scale and resolution of jets, and the energy resolution of photons. Generally, fits exploiting both shape and rate information
show increased sensitivity to experimental and modelling uncertainties compared with fits using only shape information.

\clearpage

\clearpage

\begin{table}[h!]
\centering
\caption{
Breakdown of the relative contributions to the uncertainty in the best-fit parameters of interest (i.e. either
$c_{H\widetilde{W}}$, $a_\mathrm{L}$, or $a_\mathrm{T}$) for the CP and polarisation studies.
The relative impact of each group of nuisance parameters is defined as the square root of the difference between
the squares of the total uncertainty obtained when treating all nuisance parameters as free and the uncertainty
obtained when that group of nuisance parameters is fixed to their best-fit values, normalised to the total uncertainty.
The statistical uncertainty component is determined by fixing all nuisance parameters to their best-fit values. The sum in
quadrature of the individual components may differ from the total uncertainty due to correlations between the different
groups of nuisance parameters. }
\resizebox{\columnwidth}{!}{
\begin{tabular}{lcccccc}
\toprule
Source                              & \makecell[c]{$c_{H\widetilde{W}}$\\(linear-only)}   & \makecell[c]{$c_{H\widetilde{W}}$\\(linear+quad.)} & \makecell[c]{$a_\mathrm{L}$\\(shape-only)} & \makecell[c]{$a_\mathrm{L}$\\(shape+rate)} & \makecell[c]{$a_\mathrm{T}$\\(shape-only)} & \makecell[c]{$a_\mathrm{T}$\\(shape+rate)}  \\
\midrule
Total statistical                   & $99\%$       &   $99\%$     &  $98\%$   &   $79\%$   & $98\%$    &  $89\%$      \\
\midrule
Total systematic                    &  $10\%$      &   $7.4\%$   &   $20\%$   &  $61\%$    & $20\%$     &  $45\%$     \\
~~~~Spurious Signal                 &  $12\%$      &   $11\%$    &   $9.1\%$  &  $7.4\%$   & $9.3\%$    &  $7.4\%$    \\
~~~~G4-AF3 discrepancies            &  $3.3\%$     &   $4.8\%$   &   $2.1\%$  &  $18\%$    & $3.3\%$    &  $12\%$     \\
~~~~Photons                         &  $< 1\%$     &   $< 1\%$   &   $2.9\%$  &  $22\%$    & $3.9\%$    &  $16\%$     \\
~~~~Jets                            &  $7.0\%$     &   $9.3\%$   &   $10\%$   &  $38\%$    & $12\%$     &  $24\%$     \\
~~~~Luminosity + Pile-up            &  $1.6\%$     &   $1.9\%$   &  $< 1\%$   &  $9.4\%$   & $< 1\%$    &  $7.0\%$    \\
~~~~VBF theory                      &  $5.8\%$     &   $5.4\%$   &    $8.8\%$ &  $31\%$    & $9.5\%$    &  $23\%$     \\
~~~~ggF theory                      &  $2.9\%$     &   $4.0\%$   &    $5.0\%$ &  $5.1\%$   & $6.1\%$    &  $5.9\%$    \\
~~~~Other Higgs boson theory        & $< 1\%$      &  $< 1\%$    &  $< 1\%$   &  $< 1\%$   & $< 1\%$    & $< 1\%$     \\
\midrule
\bottomrule
\end{tabular}}
\label{tab:syst_breakdown}
\end{table}

\subsection{Combination of searches for CP violating effects using Run-2 and Run-3 results} \label{sec:results_combination}

The results obtained from the Run-3 dataset are combined with the previous ATLAS search for CP-violating effects in VBF
$H\rightarrow \gamma\gamma$ events using Run-2 data~\cite{HIGG-2020-08}. To maximise the sensitivity to $c_{H\widetilde{W}}$,
a simultaneous fit to the Run-2 and Run-3 datasets is performed. Theoretical uncertainties in
the parton shower model, are treated as fully correlated across both datasets. The QCD scale uncertainty
is correlated for the ggF process but uncorrelated for the VBF signal. This accounts for the different uncertainty calculation
procedures employed in the Run-2 and Run-3 analyses. Most experimental uncertainties are treated as fully correlated,
with exceptions for photon and jet energy scale uncertainties related to specific corrections for the AF3 simulation, photon leakage effects,
and the extrapolation between Run-2 and Run-3 conditions. All other systematic uncertainties are treated as uncorrelated. The spurious signal
and parton shower uncertainties represent the dominant systematic components in the combination. The impact of the specific correlation
scheme was evaluated and found to be negligible, as the total uncertainty is dominated by the statistical component. \par

The joint likelihood is constructed as the product of the individual likelihood functions from the Run-2 and Run-3 analyses. Correlated systematic
sources share a common NP. In this combination, the normalisation scale factor $\mu_{ggF}$ is treated as a free-floating parameter in the
Run-2 component of the fit to ensure a consistent treatment across both datasets. This approach results in a $\sim5\%$ broadening of the
$c_{H\widetilde{W}}$ constraint relative to the original Run-2 publication, where $\mu_{ggF}$ was constrained to its SM prediction within
theoretical uncertainties.\par

\begin{figure}[h!]
\centering
\subfloat{\includegraphics[width=0.605\textwidth]{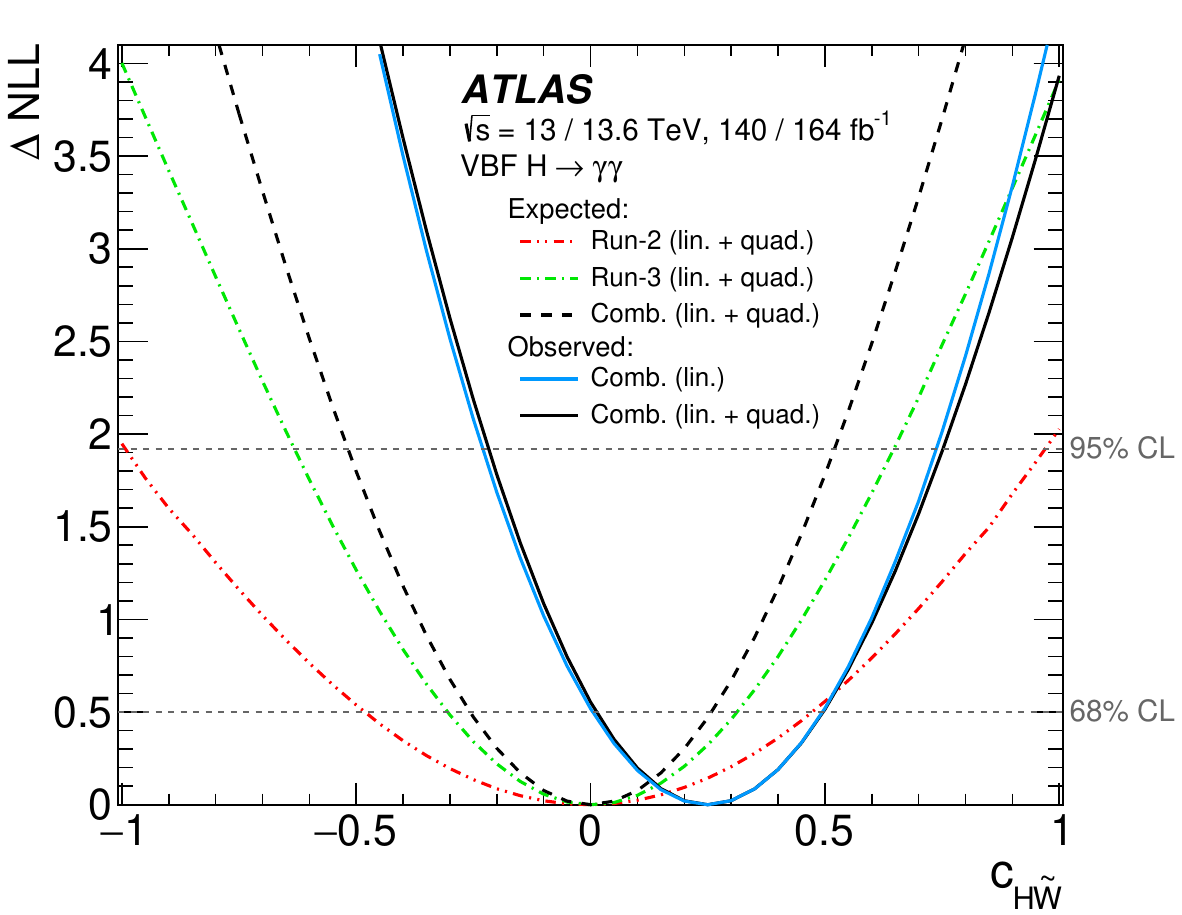}}
\caption{Expected and observed likelihood curve as a function of the Wilson coefficient $c_{H\widetilde{W}}$. The results are shown for the
individual Run-2 and Run-3 datasets, as well as for the combined fit of both datasets. Scans are performed using both the full
linear+quadratic and linear-only parameterisations. The dashed horizontal lines indicate the $\Delta$NLL values corresponding to the $68\%$ and $95\%$
CL intervals. Results are obtained assuming a new physics scale of $\Lambda = 1\,\TeV$.}
\label{fig:combination}
\end{figure}

Confidence level intervals are determined for the parameterisations with linear-only and linear+quadratic terms. The expected and observed profile likelihood scans
for the Run-2, Run-3, and combined datasets are shown in Figure~\ref{fig:combination}. For the fit with the linear+quadratic parameterisation, the best-fit value
is found to be $c_{H\widetilde{W}} = 0.25$. The observed $68\%$ ($95\%$) CL intervals for $c_{H\widetilde{W}}$ are determined to be $[0.01, 0.49]$ ($[-0.23, 0.75]$)
compared to the expected intervals of $[-0.26, 0.26]$ ($[-0.52, 0.52]$). The corresponding normalisation scale factors at the best-fit point are
$\mu_\mathrm{VBF} = 1.14 \pm 0.12$ and $\mu_\mathrm{ggF} = 1.43 ^{+0.59}_{-0.47}$. For the combined fit, the impact of the quadratic terms on the CL intervals
is negligible. The best-fit value, and the expected and observed $95\%$ CL intervals for the combined fit are also presented alongside the individual
Run-2 and Run-3 results in Figure~\ref{fig:Summary_of_results_cp}.

%

%
\section{Conclusion}\label{sec:conclusion}

%

This article presents constraints on the CP structure and the polarisation-dependent coupling strengths of the $HVV$ coupling, probed via the vector-boson fusion
production mode. The results are obtained using the $H\left(\rightarrow\gamma\gamma\right) jj$ final state in data corresponding to an integrated luminosity
of 164 fb$^{-1}$ of proton--proton collision data at $\sqrt{s} = 13.6\,\TeV$, recorded with the ATLAS detector during 2022--2024.
Significant improvements with respect to previous results are achieved through the implementation of a new neural network-based classification algorithm and the
larger analysed data sample.

The CP-odd Wilson coefficient $c_{H\widetilde{W}}$, parameterising CP-violating effects in the $HVV$ coupling, and the coupling-strength scale factors
for longitudinally and transversely polarised vector bosons, are constrained using kinematic observables. Based on the shape of an Optimal Observable,
the $68\%$ ($95\%$) CL intervals for $c_{H\widetilde{W}}$ are determined to be $[-0.06,0.54]$ ($[-0.35,0.88]$). Compared with a similar analysis
of $H(\rightarrow \gamma\gamma) jj$ events using 140\,fb$^{-1}$ of $\sqrt{s}=13\,\TeV$ data, these Run-3 constraints show a $38\%$ improvement,
with $29\%$ arising from refined analysis techniques, most notably the implementation of a new NN-based classification algorithm. To maximise sensitivity,
the Run-2 and Run-3 results are combined, narrowing the $68\%$ ($95\%$) CL intervals for $c_{H\widetilde{W}}$ to $[0.01,0.49]$ ($[-0.23,0.75]$),
which represents a total improvement of $50\%$ over the Run-2 results.

Two fit configurations are used to determine the polarisation-dependent scale factors. Using shape information from the $\Delta\Phi_{jj}$ distribution
alone, the $68\%$ ($95\%$) CL intervals are $a_\mathrm{L}\in[0.94,1.36]$ ($[0.84,2.02]$) and $a_\mathrm{T}\in[0.74,1.06]$
($[0.50,1.20]$). When both shape and rate information are used, the constraints are significantly strengthened to $a_\mathrm{L}\in[0.98,1.11]$ ($[0.93,1.17]$)
and $a_\mathrm{T}\in[0.80,1.03]$ ($[0.70,1.14]$).  These measurements represent the first determination of polarisation-dependent coupling strength
scale factors in $H\left(\rightarrow\gamma\gamma\right) jj$ events. Compared with previous results in $H(\rightarrow WW^{*}\rightarrow e\nu\mu\nu) jj$
events using 36.1\,fb$^{-1}$ of $\sqrt{s}=13\,\TeV$ data, the current analysis provides a major improvement in sensitivity. Using shape information
alone, this analysis provides $95\%$ CL intervals for $a_\mathrm{L}$ and $a_\mathrm{T}$, whereas the sensitivity of the previous analysis was insufficient
to reach those confidence levels. Furthermore, when comparing the results from fits that use both shape and rate information,
the widths of the 95\% CL intervals for $a_\mathrm{L}$ and $a_\mathrm{T}$ are reduced by more than a factor of three relative to the results
obtained in $H(\rightarrow WW^{*}\rightarrow e\nu\mu\nu) jj$ events.

All results are consistent with the SM expectations, as no significant CP-odd contributions or deviations in polarisation-dependent couplings are observed.
Notably, these studies represent a significant advancement as the first ATLAS results to rely exclusively on the AF3 detector simulation of the ATLAS calorimeter
system for all Monte Carlo samples, demonstrating the viability of this fast simulation tool.


%
\section*{Acknowledgements}

%
%

%
%

We thank CERN for the very successful operation of the LHC and its injectors, as well as the support staff at
CERN and at our institutions worldwide without whom ATLAS could not be operated efficiently.

The crucial computing support from all WLCG partners is acknowledged gratefully, in particular from CERN, the ATLAS Tier-1 facilities at TRIUMF/SFU (Canada), NDGF (Denmark, Norway, Sweden), CC-IN2P3 (France), KIT/GridKA (Germany), INFN-CNAF (Italy), NL-T1 (Netherlands), PIC (Spain), RAL (UK) and BNL (USA), the Tier-2 facilities worldwide and large non-WLCG resource providers. Major contributors of computing resources are listed in Ref.~\cite{ATL-SOFT-PUB-2026-001}.

We gratefully acknowledge the support of ANPCyT, Argentina; YerPhI, Armenia; ARC, Australia; BMWFW and FWF, Austria; ANAS, Azerbaijan; CNPq and FAPESP, Brazil; NSERC, NRC and CFI, Canada; CERN; ANID, Chile; CAS, MOST, NSFC and FRFCU, China; Minciencias, Colombia; MEYS CR, Czech Republic; DNRF and DNSRC, Denmark; IN2P3-CNRS and CEA-DRF/IRFU, France; SRNSFG, Georgia; BMFTR, HGF and MPG, Germany; GSRI, Greece; RGC and Hong Kong SAR, China; ICHEP and Academy of Sciences and Humanities, Israel; INFN, Italy; MEXT and JSPS, Japan; CNRST, Morocco; NWO, Netherlands; RCN, Norway; MNiSW, Poland; FCT, Portugal; MNE/IFA, Romania; MSTDI, Serbia; MSSR, Slovakia; ARIS and MVZI, Slovenia; DSI/NRF, South Africa; MICIU/AEI, Spain; SRC and Wallenberg Foundation, Sweden; SERI, SNSF and Cantons of Bern and Geneva, Switzerland; NSTC, Taipei; TENMAK, T\"urkiye; STFC/UKRI, United Kingdom; DOE and NSF, United States of America.

Individual groups and members have received support from BCKDF, CANARIE, CRC and DRAC, Canada; CERN-CZ, FORTE and PRIMUS, Czech Republic; COST, ERC, ERDF, Horizon 2020 and Marie Sk{\l}odowska-Curie Actions, European Union; Investissements d'Avenir Labex, Investissements d'Avenir Idex and ANR, France; DFG and AvH Foundation, Germany; Herakleitos, Thales and Aristeia programmes co-financed by EU-ESF and the Greek NSRF, Greece; BSF-NSF and MINERVA, Israel; NCN and NAWA, Poland; La Caixa Banking Foundation, CERCA and AGAUR programs from Generalitat de Catalunya and PROMETEO and GenT Programmes Generalitat Valenciana, Spain; G\"{o}ran Gustafssons Stiftelse, Sweden; The Royal Society and Leverhulme Trust, United Kingdom; Eric and Wendy Schmidt Fund for Strategic Innovation, United States of America.

In addition, individual members wish to acknowledge support from Chile: Agencia Nacional de Investigaci\'on y Desarrollo (ANID FONDECYT reg. 1230987, FONDECYT 1230812, FONDECYT 1240864, Fondecyt 3240661, Fondecyt Regular 1240721); China: Fundamental Research Funds for the Central Universities (010-63263105), Chinese Ministry of Science and Technology (MOST-2023YFA1605700, MOST-2023YFA1609300), National Natural Science Foundation of China (NSFC 12275265, NSFC-W2543005); Czech Republic: Czech Science Foundation (GACR - 24-11373S), Ministry of Education Youth and Sports (ERC-CZ-LL2327, FORTE CZ.02.01.01/00/22\_008/0004632), PRIMUS Research Programme (PRIMUS/21/SCI/017); EU: H2020 European Research Council (ERC - 101002463); European Union: European Research Council (BARD No. 101116429, ERC - 101219398, ERC - 948254, ERC 101089007), European Regional Development Fund (HE COFUND GA No.101081355, ERDF), Marie Sklodowska-Curie Actions (GAP-101168829); France: Agence Nationale de la Recherche (ANR-21-CE31-0013, ANR-22-EDIR-0002, ANR-24-CE31-0504-01); Germany: Deutsche Forschungsgemeinschaft (DFG - 469666862); China: Research Grants Council (GRF); Italy: Istituto Nazionale di Fisica Nucleare (LHC-MIUR - 28003/2025), Ministero dell'Università e della Ricerca (NextGenEU 153D23001490006 M4C2.1.1, NextGenEU I53D23000820006 M4C2.1.1, NextGenEU I53D23001490006 M4C2.1.1, SOE2024\_0000023); Japan: Japan Society for the Promotion of Science (JSPS KAKENHI  JP25H0063, JSPS KAKENHI JP22H01227, JSPS KAKENHI JP22H04944, JSPS KAKENHI JP22KK0227, JSPS KAKENHI JP24K23939, JSPS KAKENHI JP24KK0251, JSPS KAKENHI JP25H00650, JSPS KAKENHI JP25H01291, JSPS KAKENHI JP25K01011, JSPS KAKENHI JP25K01023); Poland: Polish National Science Centre (NCN 2021/42/E/ST2/00350, NCN OPUS 2023/51/B/ST2/02507, NCN OPUS nr 2022/47/B/ST2/03059, NCN UMO-2019/34/E/ST2/00393, UMO-2022/47/O/ST2/00148, UMO-2023/49/B/ST2/04085, UMO-2023/51/B/ST2/00920, UMO-2024/53/N/ST2/00869); Spain: Agència de Gestió d'Ajuts Universitaris i de Recerca. (AGAUR - 2023 BP 00141), Ministry of Science and Innovation (RYC2019-028510-I, RYC2020-030254-I, RYC2021-031273-I, RYC2022-038164-I), Ministerio de Ciencia, Innovación y Universidades/Agencia Estatal de Investigaci\'on (EU NextGenerationEU (PRTR-C17.I1), PID2022-142604OB-C22); Sweden: Carl Trygger Foundation (Carl Trygger Foundation CTS 22:2312), Swedish Research Council (Swedish Research Council 2023-04654, VR 2021-03651, VR 2022-03845, VR 2022-04683, VR 2023-03403, VR 2024-05451, VR 2025-05940), Knut and Alice Wallenberg Foundation (KAW 2023.0366); Switzerland: Swiss National Science Foundation (SNSF - PCEFP2\_194658); United Kingdom: The Binks Trust, Royal Society (NIF-R1-231091); United States of America: U.S. Department of Energy (ECA DE-AC02-76SF00515), John Templeton Foundation (John Templeton Foundation 63206), Neubauer Family Foundation.

%
%


\clearpage
\printbibliography
\clearpage
 
\begin{flushleft}
\hypersetup{urlcolor=black}
{\Large The ATLAS Collaboration}

\bigskip

\AtlasOrcid[0000-0002-6665-4934]{G.~Aad}$^\textrm{\scriptsize 102}$,
\AtlasOrcid[0000-0001-7616-1554]{E.~Aakvaag}$^\textrm{\scriptsize 17}$,
\AtlasOrcid[0000-0002-5888-2734]{B.~Abbott}$^\textrm{\scriptsize 121}$,
\AtlasOrcid[0000-0002-0287-5869]{S.~Abdelhameed}$^\textrm{\scriptsize 83b}$,
\AtlasOrcid[0000-0002-1002-1652]{K.~Abeling}$^\textrm{\scriptsize 54}$,
\AtlasOrcid[0000-0001-5763-2760]{N.J.~Abicht}$^\textrm{\scriptsize 48}$,
\AtlasOrcid[0000-0002-8496-9294]{S.H.~Abidi}$^\textrm{\scriptsize 30}$,
\AtlasOrcid[0009-0003-6578-220X]{M.~Aboelela}$^\textrm{\scriptsize 44}$,
\AtlasOrcid[0000-0002-9987-2292]{A.~Aboulhorma}$^\textrm{\scriptsize 36e}$,
\AtlasOrcid[0000-0001-5329-6640]{H.~Abramowicz}$^\textrm{\scriptsize 154}$
% ...
\bigskip
\\

% ...
$^{17}$Department for Physics and Technology, University of Bergen, Bergen; Norway.\\
% ...
$^{30}$Physics Department, Brookhaven National Laboratory, Upton NY; United States of America.\\
% ...
$^{36}$$^{(a)}$Facult\'e des Sciences Ain Chock, Universit\'e Hassan II de Casablanca;$^{(b)}$Facult\'{e} des Sciences, Universit\'{e} Ibn-Tofail, K\'{e}nitra;$^{(c)}$Facult\'e des Sciences Semlalia, Universit\'e Cadi Ayyad, LPHEA-Marrakech;$^{(d)}$LPMR, Facult\'e des Sciences, Universit\'e Mohamed Premier, Oujda;$^{(e)}$Facult\'e des sciences, Universit\'e Mohammed V, Rabat;$^{(f)}$Institute of Applied Physics, Mohammed VI Polytechnic University, Ben Guerir; Morocco.\\
% ...
$^{44}$Physics Department, Southern Methodist University, Dallas TX; United States of America.\\
% ...
$^{48}$Fakult\"{a}t Physik, Technische Universit{\"a}t Dortmund, Dortmund; Germany.\\
% ...
$^{54}$II. Physikalisches Institut, Georg-August-Universit\"{a}t G\"ottingen, G\"ottingen; Germany.\\
% ...
$^{83}$$^{(a)}$Khalifa University of Science and Technology, Abu Dhabi;$^{(b)}$New York University Abu Dhabi, Abu Dhabi;$^{(c)}$United Arab Emirates University, Al Ain;$^{(d)}$University of Sharjah, Sharjah; United Arab Emirates.\\
% ...
$^{102}$CPPM, Aix-Marseille Universit\'e, CNRS/IN2P3, Marseille; France.\\
% ...
$^{121}$Homer L. Dodge Department of Physics and Astronomy, University of Oklahoma, Norman OK; United States of America.\\
% ...
$^{154}$Raymond and Beverly Sackler School of Physics and Astronomy, Tel Aviv University, Tel Aviv; Israel.\\
% ...
$^{a}$ Also at Affiliated with an institute formerly covered by a cooperation agreement with CERN.\\
$^{b}$ Also at An-Najah National University, Nablus; Palestine.\\
$^{c}$ Also at Borough of Manhattan Community College, City University of New York, New York NY; United States of America.\\
$^{d}$ Also at Center for Interdisciplinary Research and Innovation (CIRI-AUTH), Thessaloniki; Greece.\\
$^{e}$ Also at Centre of Physics of the Universities of Minho and Porto (CF-UM-UP); Portugal.\\
$^{f}$ Also at CERN, Geneva; Switzerland.\\
% ...

\end{flushleft}


%
%

%
%
%
%

\end{document}